\documentclass[a4paper,11pt]{article}
\usepackage[utf8]{inputenc}
\usepackage[english]{babel}
\usepackage{braket}
\usepackage{xspace}
\usepackage{bbm}
\usepackage{mathdots}
\usepackage{stackrel}
\usepackage{xcolor}
\usepackage{mathtools}
\usepackage{bbm}
\usepackage{subfigure}
\usepackage[T1]{fontenc}               
\usepackage[left=3cm,
            right=2.5cm,
            top=2.5cm,
            bottom=3cm
            ]{geometry}                
\usepackage{graphicx}
\usepackage{mathrsfs}
\usepackage{appendix}
\usepackage{fancyhdr}
\usepackage{amsmath,                   
            amssymb,                   
            amsthm}                    
\usepackage{cite}

\usepackage{setspace} 

\usepackage[colorlinks=true,linkcolor=blue, citecolor=blue, bookmarks]{hyperref}

\numberwithin{equation}{section}

\def \be {\begin{equation}} 
\def \ee {\end{equation}} 
\def \l {\left(} 
\def \r {\right)} 
\def \la {\langle} 
\def \ra {\rangle}  

\date{}

\title{Full counting statistics and symmetry resolved entanglement for free conformal theories with interface defects}

\author{Luca Capizzi$^1$, Sara Murciano$^{2,3}$, and Pasquale Calabrese$^{1,4}$}
\begin{document}
\maketitle

\maketitle
{\small
\vspace{-5mm}  \ \\
{$^{1}$}  SISSA and INFN Sezione di Trieste, via Bonomea 265, 34136 Trieste, Italy\\
\medskip
$^{2}$	Walter Burke Institute for Theoretical Physics, Caltech, Pasadena, CA 91125, USA\\[-0.1cm]
\medskip
$^{3}$	Department of Physics and IQIM, Caltech, Pasadena, CA 91125, USA\\[-0.1cm]
\medskip
{$^{4}$}  International Centre for Theoretical Physics (ICTP), Strada Costiera 11, 34151 Trieste, Italy\\
\medskip
}
\begin{abstract} 

We consider the ground state of two species of one-dimensional critical free theories coupled together via a conformal interface. They have an internal $U(1)$ global symmetry and we investigate the quantum fluctuations of the total charge on one side of the interface, giving analytical predictions for the full counting statistics, the charged moments of the reduced density matrix and the symmetry resolved R\'enyi entropies. Our approach is based on the relation between the geometry with the defect and the homogeneous one, and it provides a way to characterise the spectral properties of the correlation functions restricted to one of the two species. Our analytical predictions are tested numerically, finding a perfect agreement.

\end{abstract}

\tableofcontents

\section{Introduction}
Measurements of a given observable are at the origin of the probabilistic nature of quantum mechanics: despite a system is prepared in identical conditions, its measurement can provide different outcomes. Therefore, it is interesting to study the full probability distribution function (PDF) of an observable in any quantum mechanical system, especially when the first few moments do not provide a good description of the distribution. Quantify these PDFs, or equivalently their full counting statistics
(FCS), has become of increasing relevance in one-dimensional quantum many-body systems, through exact derivations, both in and out of equilibrium, and in inhomogeneous systems (the interested readers
can consult the comprehensive literature on the subject~\cite{ll-93,s-07,cd-07,bss-07,aem-08,lp-08,ia-13,sk-13,e-13,k-14,mcsc-15,sp-17,CoEG17,nr-17,hb-17,bpc-18,gadp-06,er-13,lddz-15,cdcd-20,c-19,bp-18,ce-20,ag-19,gec-18,nr-20,ppg-19,dbdm-20,cgcm-20,vcc-20, bcckr-22,dm-20,ehm-09,bd-16}, from equilibrium setups to charge transport). This large number of theoretical works about the FCS has been also motivated by the recent groundbreaking advances with cold
atom experiments~\cite{HLSI08,KPIS10,KISD11,GKLK12,AJKB10,JABK11,zyz-22}. 
Shifting the attention to observables with support on a finite, but large, subsystem embedded
in a thermodynamic system provides a further motivation for studying the FCS: indeed, it turned out to be strictly related to the
entanglement entropy of the same subsystem~\cite{kl-09,kl-09b,hgf-09,sfr-11a,sfr-11b,cmv-12fcs,lbb-12,si-13,clm-15,u-19,arv-20,bcp-23,dvdvdr-23} and so it provides
indirect information about the latter. This interplay will be relevant in our work, too, especially for the connections between the FCS and the entanglement entropy in a given charge sector in the presence of a global symmetry. 

In order to make this connection explicit, we remind here the definition of the FCS. We consider an extended quantum system in a pure state $\ket{\Psi}$, defined in a bipartite Hilbert space $\mathcal{H}=\mathcal{H}_A\otimes \mathcal{H}_B$, where $\mathcal{H}_A$ and $ \mathcal{H}_B$ are respectively associated to two spatial regions $A$ and $B$. The reduced density matrix, $\rho_A$, obtained by taking the partial trace to the complementary subsystem 
as $\rho_A=\mathrm{Tr}_B(\ket{\Psi}\bra{\Psi})$, describes the state 
in $A$. We take as an observable a charge operator $Q$ generating a $U(1)$ symmetry group, such that it is the sum of the charge in each region, $Q=Q_A+Q_B$ (although this ultralocality condition is not required for the FCS it is fundamental for the symmetry resolved entanglement \cite{goldstein}). The FCS is defined as 
\begin{equation}\label{eq:def}
 Z_1(\alpha)=\mathrm{Tr}[\rho_A e^{i\alpha Q_A}].
\end{equation}
The FCS is the Fourier transform of the PDF of the charge $Q_A$.
We observe that in this case $Z_1(\alpha)$ is also equal to $\mathrm{Tr}[\rho \,e^{i\alpha Q_A}]$, while this equivalence breaks down when we consider
\begin{equation}\label{eq:charge_def}
    Z_n(\alpha)=\mathrm{Tr}[\rho^n_A e^{i\alpha Q_A}].
\end{equation}
The latter quantities are know as \textit{charged moments} of the reduced density matrix~\cite{Belin-Myers-13-HolChargedEnt}. These objects have played a crucial role in the study of the symmetry resolution of the entanglement~\cite{lr-14,goldstein,xavier}. As we will show in the following sections, if $\ket{\Psi}$ is an eigenstate of $Q$, then $[\rho_A, Q_A]=0$ and $\rho_A$ displays a block-diagonal structure in the charge sectors of $Q_A$, labelled by its eigenvalues $q$. Then,  the moments of $\rho_A$ restricted to the $q$ charge sector are the Fourier transform of $Z_n(\alpha)$ defined in Eq.~\eqref{eq:charge_def} \cite{goldstein,xavier}.
Also when $[\rho_A, Q_A]\neq0$, quantities analogous to \eqref{eq:charge_def} provide crucial information about the system of interest \cite{amc-22,amvc-23}.

In this manuscript, we are interested in the study of both the FCS and the charged moments of two species of massless complex
fermions or bosons in one dimension coupled together via a conformal interface. A similar analysis has been done in Refs.~\cite{cmc-22} and~\cite{cmc-22a} to compute the total R\'enyi entanglement entropies~\cite{cc-04}, 
\begin{equation}
    S_n=\frac1{1-n}\log \mathrm{Tr}\rho_A^n,
\end{equation}
and the entanglement negativity~\cite{cct-12} among an arbitrary number of fermionic and bosonic species, respectively. In both cases, we have found that these entanglement measures
show a logarithmic growth with the system size, and the universal prefactor depends
both on the details of the interface and the bipartition. This is consistent with the crucial result that, for free massless
theories, the conformal junction is equivalent to a marginal defect \cite{kf-92}, and so can alter the leading behaviour of the entanglement measures~\cite{Levine,ss-08,bb-15,bbr-13,tm-21,cmv-12,ep-12,p-05,ep-10,ep2-12,ce-22,csrc-23,rs-22,sths-22}. 
Given the success of the entanglement
entropies in the description of the universality and of the relevance of the parameters defining the junction, it is natural to wonder whether this also occurs both for the probability distribution of the charge and for the entanglement resolved in the different charge sectors. We address this problem here by studying two different kinds of free systems, which provide an ideal framework to perform exact computations both analytically and numerically.

\paragraph{Main result:} We summarise here the main results of this paper so that the reader can easily find them without dealing with all technical details of the derivation. 
For a system of free fermions with an interface parametrised by the transmission probability $\mathcal{T}$ and when the subsystem $A$ is one of the two halves of the system on one side of the defect, we show that the charged moments defined in Eq. \eqref{eq:charge_def} behave at leading order in the system size, $L$, as 
\begin{equation}
\begin{split}
    \log Z_n(\alpha)=&\log L \int^1_{0}dz \  \rho_{C'_A}(z)\log(z^ne^{i\alpha/2}+(1-z)^ne^{-i\alpha/2}),\\   
\rho_{C'_A}(z) =& \begin{cases}\frac{1}{2\pi^2}\frac{|1-2z|}{z(1-z)\sqrt{1-4z(1-z)/\mathcal{T}}}, \quad z \in (0,\frac{1-\sqrt{1-\mathcal{T}}}{2})\cup (\frac{1+\sqrt{1-\mathcal{T}}}{2},1),\\
0, \quad \text{otherwise}.
\end{cases}
\end{split}
\end{equation}
We stress that the prefactor of the logarithmic term  does depend on the details of the interface, on the replica index $n$ and on the flux $\alpha$. 
Similarly, for a free bosonic system we find  
\begin{equation}
\begin{split}
    \log Z_n(\alpha)=&\log L \int^\infty_{1/4}
 dz \rho_{X'_A P'_A}(z)\log((\sqrt{z}-1/2)^{2n}+(\sqrt{z}+1/2)^{2n}-2\cos \alpha(z-1/4)^{n}),\\   
\rho_{X'_A P'_A}(z) =& \begin{cases}\frac{1}{2\pi^2}\frac{1}{(z-1/4)\sqrt{1+(4z-1)/\mathcal{T}}}, \quad z \in (1/4,\infty),\\
0, \quad \text{otherwise},
\end{cases}
\end{split}
\end{equation}
where again the prefactor of the logarithmic term does depend on $\mathcal{T}, n ,\alpha$, but in a different way with respect to the fermionic case. 
These results generalise what obtained in Ref. \cite{cmc-22a} for $\alpha=0$ and in Ref. \cite{mgc-20} in the absence of the defect (i.e. ${\cal T}=0$ and arbitrary $\alpha$). 
Beyond the originality of the results, we also stress that the techniques we use here is based on the evaluation of the spectral density of the correlation matrix ($\rho_{X'_A P'_A}(z)$ for bosons and $\rho_{C'_A}$ for fermions) which represents a further new result of our work.   
When $n=1$, $Z_1(\alpha)$ is the FCS in Eq. \eqref{eq:def}. Even though the leading order behaviour of the charged moments can be derived using different and easier approaches, as we will show in the main text (especially for $n=1$), we stress that the merit of our procedure is that it immediately gives the result for any real value of $n$, without the need of any complicated analytic continuation. This allows us to straightforwardly evaluate the symmetry resolution of the entanglement, which is the second main result of this manuscript. In this respect, we show that, at leading order in $L$, the symmetry resolved entanglement is equally distributed among the different charge sectors, and the presence of the interface does not affect the \textit{equipartition of the entanglement} \cite{xavier}. 
\paragraph{Outline:} The structure of the manuscript is the following. In Sec.~\ref{sec:fermions}, we consider a free fermion gas in which the presence of the interface introduces a marginal perturbation. The interface boundary conditions are defined by a specific scattering matrix $S$, and we can exploit the gaussianity of the model to compute both the FCS and the charged moments through the correlation matrix of the system and, especially, its spectral density. This approach immediately provides a result for the charged moments in Eq.~\eqref{eq:charge_def} for any non-integer $n$ index and, as a byproduct of our findings, we can also study the symmetry resolution of the entanglement in this setup. In Sec.~\ref{sec:boson}, we  tackle the problem of two coupled complex massless Klein-Gordon field theories. By using the same techniques adopted in the fermionic case, we are able to evaluate the spectral density of the correlation matrix and consequently our main quantities of interest, i.e. the FCS, the charged moments, and, eventually, the symmetry resolved entanglement. In both cases, we validate our predictions with exact lattice computations. Finally, we draw our conclusions in Sec.~\ref{sec:conclusions} and we include two appendices about the technical details of our results.

\section{Free fermions}\label{sec:fermions}
We start our analysis about the FCS of the charge operator $Q_A$ and the charged moments from the fermionic case. Closely following Ref.~\cite{cmv-12,cmc-22}, we first review the construction of the model and its correlation matrix. Then, we report the explicit computation for the spectral density. We conclude the section with the study of the symmetry resolution of the entanglement.

\subsection{The Schroedinger junction}
In this section, we analyse a fermion gas on a junction made up of $2$ wires of length $L$, joined together through a single defect described by a non-trivial scattering matrix.
Each point of the junction is parametrised by a pair
\be
(x,j), \quad x \in [0,L], \quad j = 1, 2,
\ee
where $j$ is the index identifying the wire and $x$ the spatial coordinate along the wire. The bulk hamiltonian of the system is
\be\label{eq:Ham}
H = \sum^{2}_{j=1} \int^L_{0} dx\frac{1}{2}\l \partial_x \Psi^\dagger_j(x)\r\l \partial_x \Psi_j(x)\r,
\ee
with $\Psi_j,\Psi_j^\dagger$ being the fermionic fields associated to the $j$-th wire (also called Schroedinger field, from which the name Schroedinger junction). 
We consider a scattering matrix
\be
S = \begin{pmatrix} \sqrt{1-\mathcal{T}} & \sqrt{\mathcal{T}} \\ \sqrt{\mathcal{T}} & -\sqrt{1-\mathcal{T}}\end{pmatrix},
\label{eq:SMatrixRenyi}
\ee
describing the defect at $x=0$, where $\mathcal{T}$ plays the role of a transmission probability. When $\mathcal{T}=1$, the junction becomes completely transmissive, while when $\mathcal{T}=0$, it is a reflective boundary. The explicit relation between the scattering matrix $S$ and the boundary conditions for the fields at $x=0$ is
\begin{equation}
\lambda(1-S)\Psi(0)-i(1+S)\partial_x\Psi(0)=0,
\end{equation}
where $\Psi=\{\Psi_j\}_{j=1,2}$ and $\lambda$ is an arbitrary real parameter with the dimension of mass (we refer to Ref. \cite{cmv-12} for further details).

To completely specify the problem, we have to specify also the boundary conditions at $x=L$, that we choose to be Dirichlet, namely
\be
\Psi_j(L) =0, \quad j=1,2.
\ee
A convenient strategy to tackle this system is the introduction of a set of uncoupled nonphysical fields $\varphi_j(x)$~\cite{bm-06,bms-07}. For instance, we first diagonalise $S$ through a unitary transformation $\mathcal{U}$, and its eigenvalues are just $\pm 1$. Then, we define the unphysical fields $\{\varphi_j(x)\}$ as
\be\label{eq:unphys}
\Psi_i(x)= \sum_{j=1}^2\mathcal{U}_{ij}\varphi_j(x).
\ee
The boundary conditions of these new fields at the defect can be
\be
\partial_x\varphi_j(0) = 0, \quad \varphi_j(0) = 0,
\ee
hence they are either Neumann or Dirichlet, if the eigenvalue of the matrix $S$ is $1$ or $-1$, respectively.

From now on, we focus on the ground state of such system with $2N$ particles, and we repeat the main steps done in 
\cite{cmc-22} in order to find a finite-dimensional representation of the correlation functions. This allows us to compute efficiently the main quantities of interest in this manuscript, e.g. the FCS, or more generally, the charged moments.
The correlation functions of the two nonphysical fields $\varphi_j(x)$ are just the ones of a single Fermi gas made by $N$ particles, with boundary condition Neumann-Dirichlet (ND) or Dirichlet-Dirichlet (DD) at the two points $x=0,L$ respectively. Going back to the original fields, one eventually expresses the correlation function of Eq.~\eqref{eq:Ham} as 
\be
C'_{ij}(x,y) \equiv \la \Psi_j^\dagger(x)\Psi_i(y) \ra = \l \frac{1+S}{2}\r_{ij} C_{ND}(x,y) + \l \frac{1-S}{2}\r_{ij}C_{DD}(x,y),
\label{eq_Kernel}
\ee
with
\be
\begin{split}
&C_{DD}(x,y) = \frac{\sin \frac{N+1/2}{ L}\pi(x-y)}{2L\sin \frac{\pi(x-y)}{2L}} -(y\rightarrow -y)\\
&C_{ND}(x,y) =  \frac{\sin \frac{N}{ L}\pi(x-y)}{2L\sin \frac{\pi(x-y)}{2L}} + (y\rightarrow -y).
\end{split}
\ee
Here, we use the symbol $C'$ for the correlation function at transmission probability $\mathcal{T}$, and we denote with $C$ the one obtained for $\mathcal{T} =1$ (in the absence of the defect).

We have reported the result for the correlation function~\eqref{eq_Kernel} as continuous kernel of the spatial variables. It is more useful to work with a finite-dimensional representation of such correlations, which is equivalent to the 
overlap matrix approach~\cite{cmv-11}. The main result is the matrix representation of the kernel as
\be
C' =\frac{1-S}{2} \otimes \begin{pmatrix} 1 & Q\\ 0 & 0\end{pmatrix} + \frac{1+S}{2} \otimes \begin{pmatrix} 0 & 0\\ Q^\dagger & 1\end{pmatrix},
\label{eq_MatrixC}
\ee
with $Q$ being a $N\times N$ matrix defined by 
\be
Q_{n,n'}   = \frac{2 n}{\pi\l n^2-(n'-1/2)^2\r}.
\label{eq_MatrixQ}
\ee 
A detailed derivation for Eq. \eqref{eq_MatrixC} can be found in \cite{cmc-22} 
 and in the appendix \ref{app:proof1}.
This result has a very convenient form for the restriction of $C'$ to the subsystem $A$ given by the first wire, denoted here by $C'_A$. For instance, if $A$ only consists of the first wire, it is sufficient to restrict the correlation matrix over its element $S_{11}$ 
and one finally gets $C'_A$ from~\eqref{eq_MatrixC} as a $2N\times 2N$ matrix, which explicitly reads
\begin{equation}\label{eq:CA}
C'_A =\frac{1-\sqrt{1-\mathcal{T}}}{2}  \begin{pmatrix} 1 & Q\\ 0 & 0\end{pmatrix} + \frac{1+\sqrt{1-\mathcal{T}}}{2}  \begin{pmatrix} 0 & 0\\ Q^\dagger & 1\end{pmatrix}.
\end{equation}
If we restrict $A$ to the second wire, we could find a similar result.

The corresponding charged moments defined in Eq.~\eqref{eq:charge_def} can be computed from $C'_A$ using standard techniques of Gaussian states, and the final result is \cite{goldstein}
\be\label{eq:Ch_Mom_Ferm}
Z'_n(\alpha)=\text{det}\l (C'_A)^n e^{i\alpha/2} + (1-C'_A)^n e^{-i\alpha/2}\r. 
\ee
For $n=1$, we recover the FCS, which can be also conveniently written as 
\be\label{eq:Z1_hom}
\begin{split}
Z'_1(\alpha) = \l Z'_1(\alpha) Z'_1(-\alpha)\r^{1/2} = \text{det}^{1/2}\l 1-4C'_A(1-C'_A)\sin^2 (\alpha/2)\r,
\end{split}
\ee
using the property $Z'_1(\alpha) = Z'_1(-\alpha)$, which comes from the particle-hole symmetry of the model.
A well known result is that the leading behavior of the FCS  for $\mathcal{T}=1$, i.e. in the absence of the defect, is~\cite{cmv-12a,kl-09,riccarda}
\be\label{eq:FCS_Ferm}
-\log Z_1(\alpha) = \l \frac{\alpha}{2\pi}\r^2 \log N + O(1), \quad \alpha \in [-\pi,\pi],
\ee
valid in the limit $N \gg 1$, and it is periodic under $\alpha \rightarrow \alpha+2\pi$.
The expression above is the starting point of our analysis, as it will be helpful to relate it with other quantities in the presence of the defect.

\subsection{Spectral density}\label{sec:spectral_fermi}

Here, we investigate the spectral density of $C'_A$, that is its density of eigenvalues, a crucial quantity from which we can eventually extract the charged moments through Eq.~\eqref{eq:Z1_hom}.\\
We first introduce the matrices
\be\label{eq:EACA}
E_A = C_A(1-C_A), \quad E'_A = C'_A(1-C'_A),
\ee
where $C_A$ ($C'_A$) refers to the absence, $\mathcal{T}=1$, (presence, $\mathcal{T}<1$) of the defect. Since the correlation matrices satisfy
\be
0\leq C_A\leq 1, \quad 0\leq C'_A\leq 1,
\ee
the corresponding $E_A,E_A'$ matrices are bounded as 
\be
0\leq E_A \leq 1/4, \quad 0\leq E'_A \leq 1/4,
\ee
namely their eigenvalues belong to $[0,1/4]$. A fundamental relation between $E_A$ and $E'_A$, which comes directly from the definition in Eq.~\eqref{eq_MatrixC} (see Refs.~\cite{cmc-22,cmv-12} for the proof and also the appendix \ref{app:proof2}), is
\be\label{eq:trEA}
 E'_A = \mathcal{T} E_A.
\ee
The spectral properties of these matrices are conveniently encoded in the resolvent, defined as 
\be
G_{E_A}(y) \equiv \text{Tr}\l \frac{1}{y-E_A} - \frac{1}{y} \r, \quad G_{E'_A}(y) \equiv \text{Tr}\l \frac{1}{y-E'_A} - \frac{1}{y} \r,
\ee
where we have subtracted the term $1/y$ to discard explicitly the contributions coming from $y=0$. This term does not contribute to the universal terms of the full-counting statistics (see Eq. \eqref{eq:Z1_hom}) or the R\'enyi entropies we are interested in (see also \cite{riccarda} for the discussion about the non-universal term related to $\braket{Q_A}$ without defects).
Indeed, we can express the density of eigenvalues as
\be
\begin{split}
\text{Tr}\l \delta(y-E_A)-\delta(y)\r = \frac{1}{2\pi}\text{Im}\l G_{E_A}(y-i0^+) - G_{E_A}(y+i0^+)\r,\\
\text{Tr}\l\delta(y-E'_A)-\delta(y)\r = \frac{1}{2\pi}\text{Im}\l G_{E'_A}(y-i0^+) - G_{E'_A}(y+i0^+)\r.
\end{split}
\ee
For the sake of convenience, we define the spectral density as
\be
\begin{split}
\rho_{E_A}(y) \equiv \frac{1}{\log N}\frac{1}{2\pi}\text{Im}\l G_{E_A}(y-i0^+) - G_{E_A}(y+i0^+)\r,\\
\rho_{E'_A}(y) \equiv \frac{1}{\log N}\frac{1}{2\pi}\text{Im}\l G_{E'_A}(y-i0^+) - G_{E'_A}(y+i0^+)\r,
\end{split}
\ee
i.e. we inserted an additional normalization factor $\log N$ in the denominator to ensure a finite limit for $N\gg 1$.

\begin{figure}[htb]
\center
\includegraphics[width=0.495\textwidth]{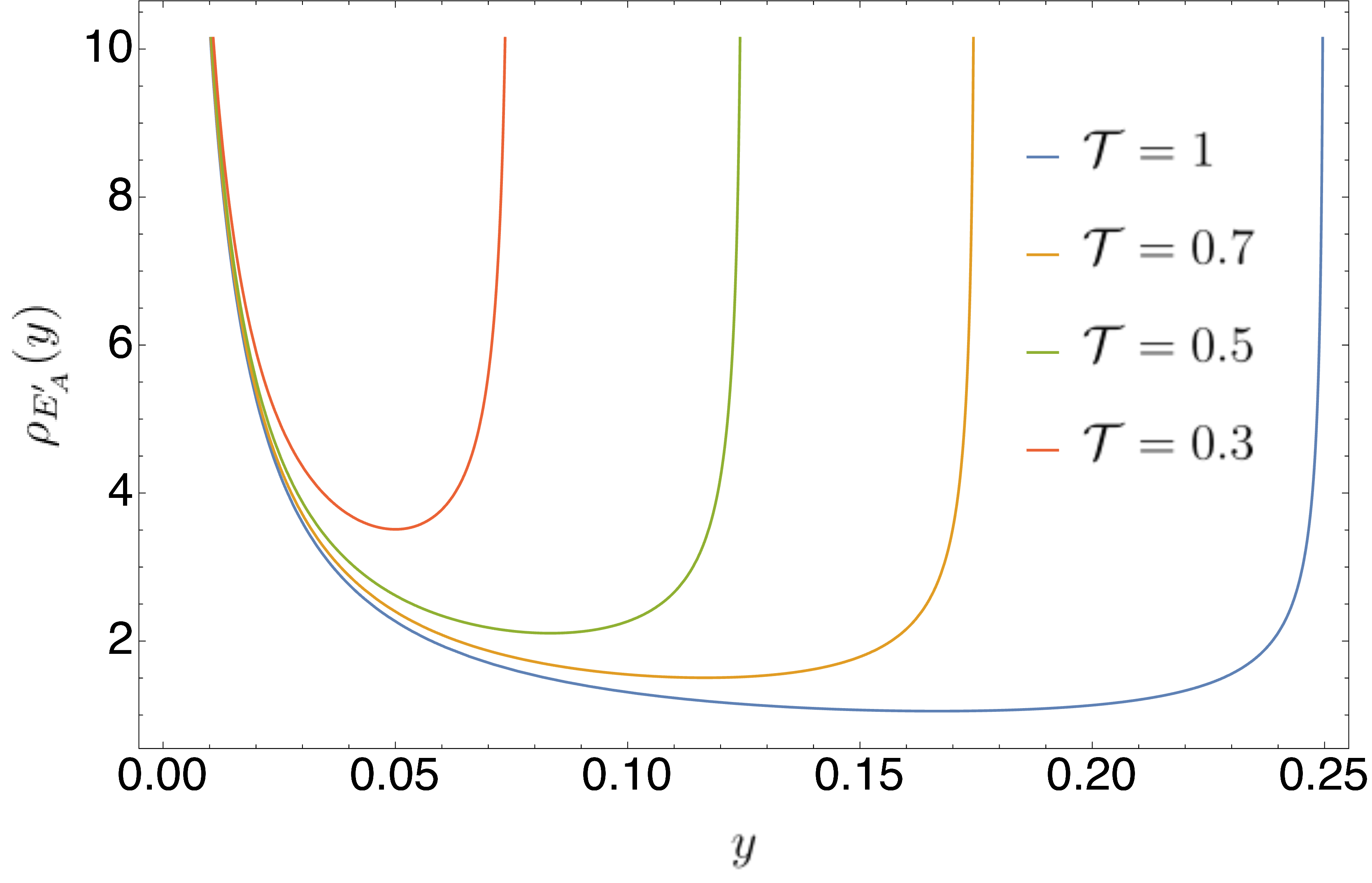}
\includegraphics[width=0.485\textwidth]{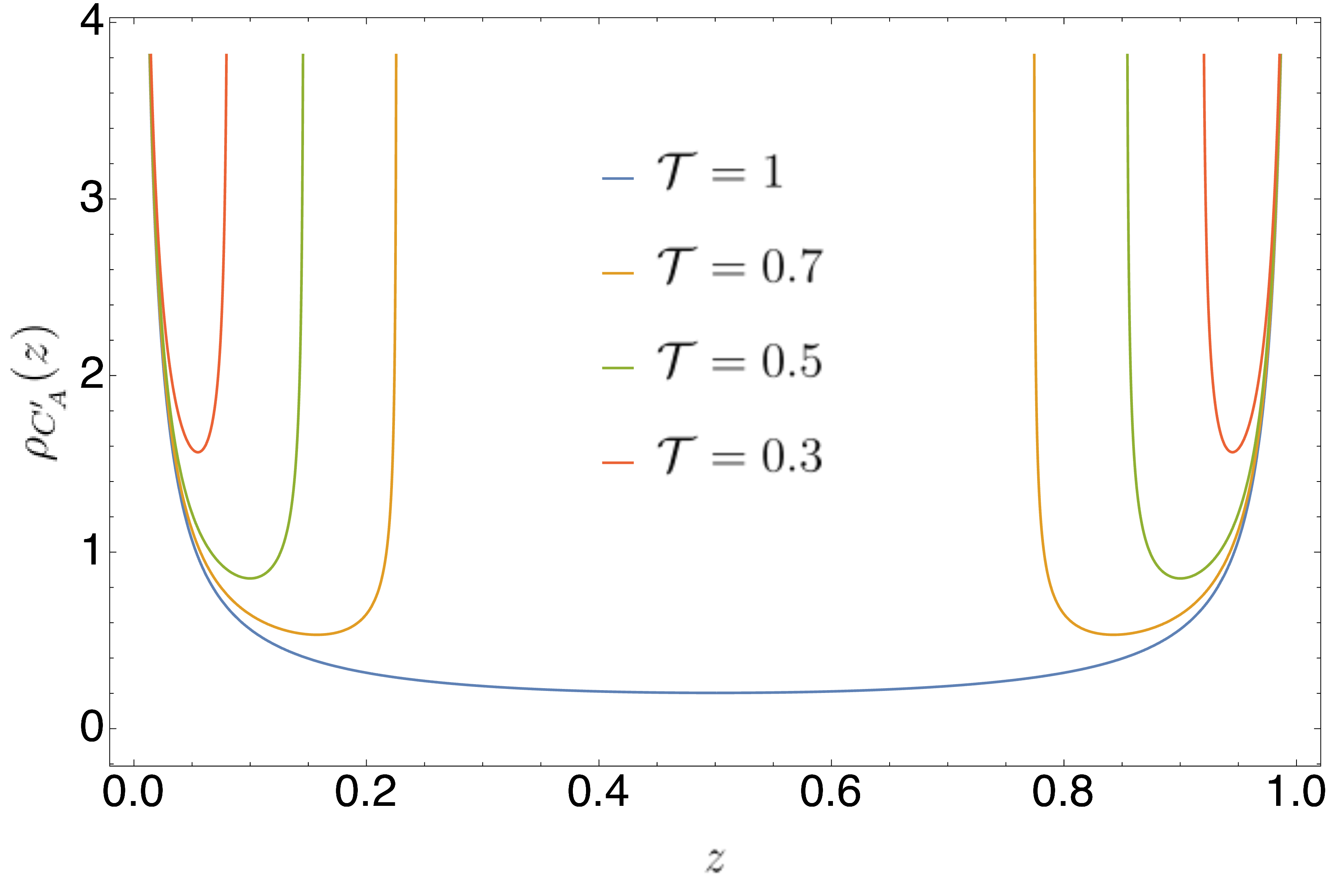}
\caption{The spectral densities of $E'_A$ (left panel) and $C'_A$ (right panel) as functions of the spectral parameter, for different values of the transmission probability $\mathcal{T}$. They correspond to the Eqs.~\eqref{eq:rhoEAp} and~\eqref{eq:rhoCp}, respectively.}
\label{fig:Rho_Ferm}
\end{figure}

In order to compute the spectral density of $C'_A$, we adopt the following procedure. We first compute $G_{E_A}(y)$ from the FCS $Z_1(\alpha)$, and we obtain the spectral density for $E_A$. Then, through Eq.~\eqref{eq:trEA}, we get the spectral density of $E'_A$, eventually related to the one of $C_A'$ by Eq.~\eqref{eq:EACA}. We start with an expression for the resolvent, obtained from Eq.~\eqref{eq:Z1_hom} with straightforward algebra, that is
\be\label{eq:Gfermions}
G_{E_A}(y) = 2\frac{d}{dy}\log Z_1(\alpha(y)),
\ee
where the relation between $\alpha$ and $y$ is given by $
y = \frac{1}{4\sin^2 (\alpha/2)}$. Inserting Eq.~\eqref{eq:FCS_Ferm} in the previous equation, we obtain the resolvent in the absence of the defect as
\be
\frac{G_{E_A}(y)}{\log N} = \frac{1}{y^{3/2}}\frac{\text{arcsin} \frac{1}{2\sqrt{y}}}{\pi^2 \sqrt{1-1/4y}}.
\ee
The expression above shows explicitly the presence of a branch-cut along $y \in [0,1/4]$, that is the support of $E_A$, which is the signal of a continuous spectral density. By using the relations
\be
\sqrt{1-\frac{1}{4y\mp i 0^+}} = \mp i \sqrt{\frac{1}{4y}-1}, \quad  \text{Re}\left[\text{arcsin}\l \frac{1}{2\sqrt{y\mp i0^+}}\r\right] = \frac{\pi}{2},
\ee
valid for $y \in (0,1/4)$, we express the spectral density as
\be\label{eq:rhoEAp}
\rho_{E_A}(y) = \frac{1}{\pi^2 } \frac{1}{y\sqrt{1-4y}}, \quad y \in [0,1/4],
\ee
while it vanishes elsewhere. For $\mathcal{T}=1$, we plot this function in the left panel of Fig.~\ref{fig:Rho_Ferm}.
We now consider $\mathcal{T}<1$, i.e. the presence of the defect, and using Eq.~\eqref{eq:trEA} we get
\be
G_{E'_A}(y) = G_{E_A}(y/\mathcal{T}), \quad \rho_{E'_A}(y) = \frac{1}{\mathcal{T}}\rho_{E_A}(y/\mathcal{T}).
\ee
In particular, it is worth to observe that the support of spectral density with the defect is shrunk to $y\in [0,\mathcal{T}/4]$ (see the left panel of Fig.~\ref{fig:Rho_Ferm}).

The last step is the computation of the spectral density of $C'_A$, from the one of $E'_A$, as a consequence of Eq.~\eqref{eq:EACA}. Here, we have to be careful as the relation between the two matrices is not invertible and two distinct eigenvalues of $C'_A$ correspond to a single eigenvalue of $E'_A$. Thus, we first focus on the region $z<1/2$, and the transformation law of densities gives 
\be\label{eq:zless12}
\rho_{C'_A}(z) = \frac{1}{2}\l \frac{dy}{dz}\r \rho_{E'_A}(y(z)),
\ee
where the relation between $z$ and $y$ is
\be
y = z(1-z).
\ee
Moreover, due to the particle-hole symmetry, namely $C'_A\rightarrow 1-C'_A$, we
have that $\rho_{C'_A}(z)$ is symmetric around $z=1/2$, that is
\be
\rho_{C'_A}(z) = \rho_{C'_A}(1-z),
\ee
and from Eq.~\eqref{eq:zless12} we obtain the result also for $z\geq 1/2$. 
We now put everything together, and we get
\be\label{eq:rhoCp}
\rho_{C'_A}(z) = \begin{cases}\frac{1}{2\pi^2}\frac{|1-2z|}{z(1-z)\sqrt{1-4z(1-z)/\mathcal{T}}}, \quad z \in (0,\frac{1-\sqrt{1-\mathcal{T}}}{2})\cup (\frac{1+\sqrt{1-\mathcal{T}}}{2},1),\\
0, \quad \text{otherwise}.
\end{cases}
\ee
A remarkable observation is that $\rho_{C'_A}(z)$ has a spectral gap for $\mathcal{T}<1$, while, as $\mathcal{T}$ approaches $1$ the gap closes, and the resulting spectral measure becomes
\be
\rho_{C_A}(z) = \frac{1}{2\pi^2 z(1-z)}, \quad z\in(0,1).
\ee
This mechanism is depicted in the right panel of Fig.~\ref{fig:Rho_Ferm}, which shows the analytical predictions for the spectral densities of $C'_A$ and $E'_A$ for different values $\mathcal{T}$.

We mention that our  finding in Eq.~\eqref{eq:rhoCp} is not completely new because a similar result was already provided in a slightly different context in Ref.~\cite{kl-09}. We also remark that our results are expected to be valid only in the thermodynamic limit $N\rightarrow \infty$; for instance, for any finite $N$, the matrix $C'_A$ has $2N$ eigenvalues and its spectral density would be given by a sum of $2N$ delta-functions centered around the eigenvalues.

Now that we have an explicit expression for the spectral density $\rho_{C'_A}(z)$, we can compute every function which depends on the spectrum of $C'_A$. In particular, we are interested in the charged moments (see Eq.~\eqref{eq:Ch_Mom_Ferm}), which read
\be\label{eq:FCSn}
-\frac{\log Z'_n(\alpha)}{\log N} = -\int^1_{0}dz \  \rho_{C'_A}(z)\log(z^ne^{i\alpha/2}+(1-z)^ne^{-i\alpha/2}),
\ee
a relation which holds for any value of $n$ (even non-integer). As a byproduct, for $n=1$ and $\mathcal{T}=1$, the integral~\eqref{eq:FCSn} gives $ \l \frac{\alpha}{2\pi}\r^2$, in agreement with Eq.~\eqref{eq:FCS_Ferm}. On the other hand, when $\mathcal{T}<1$, we find
\be\label{eq:FCST}
-\log Z'_1(\alpha) = \l \frac{\text{arcsin}(\sqrt{\mathcal{T}} \sin(\alpha/2))}{\pi}\r^2 \log N + O(1),
\ee
which is the leading term of the FCS in the presence of a defect. Notice that the same result can be obtained in a much simpler way, using the relation in Eq.~\eqref{eq:trEA}, which amounts to the replacement
\be
\sin^2(\alpha/2) \rightarrow \mathcal{T}\sin^2(\alpha/2),
\ee
in Eq.~\eqref{eq:FCS_Ferm}. We mention that the prediction in Eq.~\eqref{eq:FCST} is compatible with previous results already obtained by conformal field theory methods in Ref.~\cite{cmc-22}. However, this trick does not generalise to $n\neq 1$ and the derivation presented above is necessary.

\begin{figure}[t]
\centering
	\includegraphics[width=0.5\linewidth]{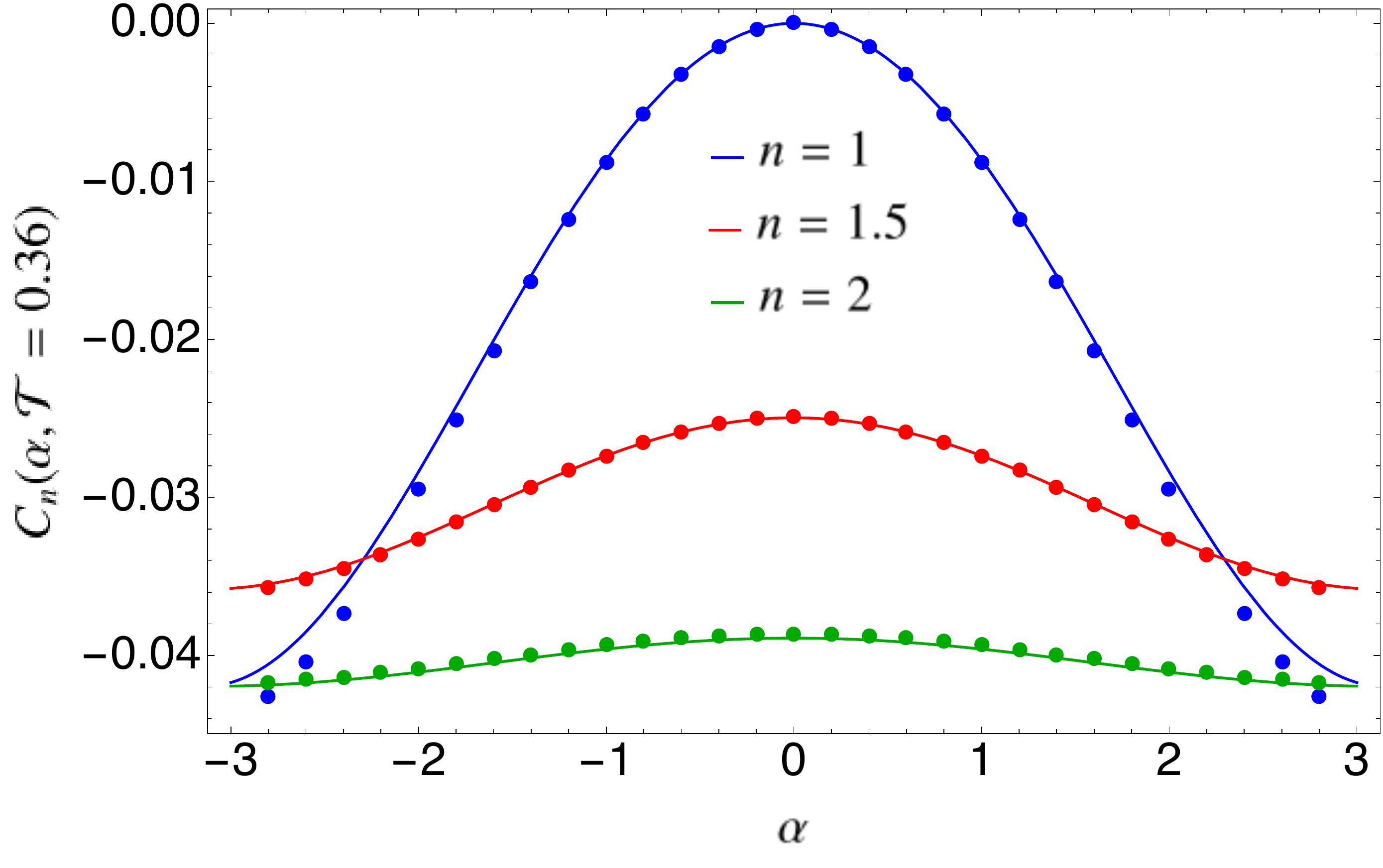}
    \caption{The coefficient of the charged moments between two wires ($A$ and $B$) as a function of $\alpha$, with fixed $\mathcal{T}=0.36$. 
    The solid line corresponds to Eq.~\eqref{eq:FCSn} while the points have been obtained through a fit of the numerics with the form $a \log N+b_0+b_1 N^{-1}$.}
    \label{fig:n1}
\end{figure}

In Fig.~\ref{fig:n1} we benchmark the coefficient of the logarithmic term of the charged moments as follows. We fix $\mathcal{T}=0.36$, and for any $\alpha \in [-\pi,\pi]$ we compute numerically the charged moments, for several values of $N$ up to 200. 
We fit the obtained numerical results with $a \log N+b_0+b_1 N^{-1}$. 
Fig.~\ref{fig:n1} reports the best fit of $a$ as a function of $\alpha$ and compares it to the corresponding analytic result in Eq.~\eqref{eq:FCSn}, finding perfect agreement.
We verify the agreement for any real value of $n$, since Eq.~\eqref{eq:FCSn} provides an analytic continuation of the charged moments in $n$. We also observe that the agreement between the numerics and our theoretical prediction worsens for $\alpha \to \pm \pi$, as it also occurs in the absence of the defect~\cite{riccarda}.

\subsubsection{Alternative derivation of $Z_n(\alpha)$ for integer $n$}
While in the derivation above we have never assumed $n$ to be an integer and
the calculation is valid for $n$ real, a simpler strategy can be employed when $n$ is integer. Indeed, thanks to the identity
\be
(x^n e^{i\alpha/2}+y^n e^{-i\alpha/2}) = \prod^{(n-1)/2}_{p=-(n-1)/2}(xe^{\frac{i}{2}(\alpha/n+2\pi p/n)}+y e^{-\frac{i}{2}(\alpha/n+2\pi p/n)}),
\ee
it is possible to show from the definition in Eq.~\eqref{eq:Ch_Mom_Ferm} that (see also~\cite{mgc-20,fmc-22})
\be\label{eq:prodfermions}
Z'_n(\alpha) = \prod^{(n-1)/2}_{p=-(n-1)/2}Z_1(2\pi p/n + \alpha/n),
\ee
where the product is over the integers/semi-integers $p$ for $n$ being odd/even. 
We mention that the same relation at $\alpha = 0$ has been employed in Ref.~\cite{cmc-22} to compute the (uncharged) moments of the reduced density matrix $\rho_A$, and here we provide a natural generalisation.
Inserting our prediction~\eqref{eq:FCST} in the previous expression, we finally get
\be\label{eq:integer}
-\log Z'_n(\alpha) = \sum^{(n-1)/2}_{p=-(n-1)/2} \l \frac{\text{arcsin}(\sqrt{\mathcal{T}} \sin(\alpha/2n +\pi p/n))}{\pi}\r^2 \log N.
\ee
This interesting result gives a relatively simple expression of the charged moments with integer $n$, while it does not immediately provide an integral representation for generic $n$ as  Eq.~\eqref{eq:FCSn} does.

\subsection{Symmetry resolved entropies}
As we already mentioned in the introduction, a very recent research line in many-body quantum systems is to understand how the entanglement organises into the various symmetry sectors of 
a theory~\cite{goldstein}. 
In the presence of a global conserved charge $Q$, the reduced density matrix can be decomposed as $\rho_A=\oplus_q P(q)\rho_{A}(q)$ in the eigenbasis of the charge restricted to the subsystem $A$, $Q_A$, and $P(q)$ is the probability of finding $q$ as an outcome of the measurement of $Q_A$ (i.e. $P(q)$ is the Fourier transform of the FCS). 
The knowledge of the charged moments allows us to access the symmetry resolved moments as
\be\label{eq:FT}
{\cal Z}_n(q)\equiv \mathrm{Tr}(\Pi_q\rho_A^n)=\displaystyle \int_{-\pi}^{\pi}\frac{d\alpha}{2\pi}e^{-i\alpha q}\mathrm{Tr}(\rho_A^n e^{i\alpha Q_A}),
\ee
where $\Pi_q$ is the projector in the charge $q$ sector, which are the eigenvalues of $Q_A$.
The symmetry resolved entropies are defined as
\begin{equation}
\label{eq:RSREE}
S_{n}(q) \equiv \dfrac{1}{1-n}\ln \mathrm{Tr} \rho^n_A(q)=  \dfrac{1}{1-n}\ln \mathrm{Tr} \frac{{\cal Z}_n(q)}{{\cal Z}_1^n(q)}.
\end{equation}
Remarkably, the symmetry resolved entanglement has been recently experimentally accessed in trapped-ion setups \cite{vek-21,ncv-21,rvm-22}. 

In order to evaluate $S_n(q)$, we expand the charged moments near $\alpha =0$ as follows
\be
-\log Z'_n(\alpha) = (A_n(\mathcal{T}) + B_n(\mathcal{T})\alpha^2 + O(\alpha^4))\log N +\dots,
\ee
with $A_n(\mathcal{T})$ and $B_n(\mathcal{T})$ being universal positive constants depending on $n$ and $\mathcal{T}$ only. The explicit expressions of those constants are
\be
A_n(\mathcal{T}) = \int^1_{0}dz \  \rho_{C'_A}(z)\log(z^n+(1-z)^n),
\ee
and
\be
B_n(\mathcal{T}) =  \int^1_{0}dz \  \rho_{C'_A}(z)\frac{z^n(1-z)^n}{2(z^n+(1-z)^n)^2},
\ee
(alternatively simpler formulae for integer $n$ can be obtained by expanding in series of $\alpha$ Eq. \eqref{eq:integer}).
By plugging this result in Eq.~\eqref{eq:FT} and performing the Fourier transform, whose corresponding integral is localized near to $\alpha \simeq 0$, we obtain 
\begin{align}\label{eq:symmF}
{\cal Z}_n(q)=& Z_n(0)\frac{e^{-\frac{q^2}{4B_n(\mathcal{T})\log N}}}{2\sqrt{\pi B_n(\mathcal{T})\log N }},\\
S_n(q) =& S_n -\frac{1}{2}\log \log 
N +
\frac{1}{4(n-1)}
 \left[\frac{1}{B_n(\mathcal{T})}
 -\frac{n}{B_1(\mathcal{T}) }\right]\frac{q^2}{\log N}+\\&\frac{1}{1-n}\log\left[(2\sqrt{\pi})^{n-1}
\frac{B_1(\mathcal{T}) ^{n/2}}
{B_n(\mathcal{T}) ^{1/2}}\right]+o(1/\log N).
\end{align}
This analysis shows that, whenever $q$ is kept fixed in the limit $N\rightarrow \infty$, the non-vanising terms are independent of $q$, a scenario which has been dubbed equipartition of entanglement~\cite{xavier}. Indeed, the first non-trivial correction to equipartition appears as a contribution of order $\sim 
q^2/\log N$ (a similar correction can be found in a different context in~\cite{longrange}).

\section{Free bosons}\label{sec:boson}

In this section, we consider a junction obtained by two wires, each one hosting a complex bosonic field. We follow closely Ref.~\cite{cmc-22a}, giving a brief review of the model and the employed techniques. Then, the evaluation of the correlation matrix and of its spectral density provides the necessary tools for the computation of the FCS and the charged moments.

\subsection{Complex massless Klein-Gordon fields}
We consider two species of bosonic fields, corresponding to the two wires, living on the segment $x \in [0,L]$ and described by the bulk hamiltonian
\be\label{eq:hcc}
H = \sum^{2}_{j=1}\int^L_0 dx \  \Pi_j^{\dagger}(x)\Pi_j(x)+ \l\partial_x \Phi_j^{\dagger}(x)\r\l\partial_x \Phi_j(x)\r,
\ee
with $\Pi_j(x), \Pi_j^{\dagger}(x)$ being the conjugated momentum associated to the bosonic fields $\Phi_j(x), \Phi_j^{\dagger}(x)$ respectively. The boundary conditions are chosen to be Dirichlet at $x=L$, namely
\be
\Phi_j(L) = \Pi_j(L)=0,
\ee
while the scale-invariant scattering matrix $S$ in Eq.~\eqref{eq:SMatrixRenyi} describes the mixing of the two species at $x=0$. We regularize the model considered above via a lattice discretization, to cure properly the UV divergences of the entanglement. For instance, we consider a chain of $N$ consecutive sites ($x=1,\dots,N$) with a lattice spacing $\varepsilon =1$, playing the role of UV cutoff, and we require
\be
\frac{L}{\varepsilon} = N+1.
\ee
In this way, one can show that Eq.~\eqref{eq:hcc} admits a discretisation in terms of a complex harmonic chain with nearest
neighbor interactions (see Refs.~\cite{altland,cmc-22a,cct-13} for more details).
By doing a change of basis which diagonalises $S$ (which has eigenvalues $\pm 1$), as in Eq.~\eqref{eq:unphys}, we obtain two species of (nonphysical) decoupled fields with either Neumann or Dirichlet boundary conditions at the defect. 

Thus, it is convenient to introduce two pair kernels, corresponding to these two choices of boundary conditions at the edges of the chain, as
\be
\begin{split}
X_{DD/ND}(x,x') \equiv&  \sum_{k=1}^N \frac{\phi^{DD/ND}_k(x)\phi^{DD/ND}_k(x')}{2\omega_k^{DD/ND}}, \\
\quad P_{DD/ND}(x,x') \equiv&   \sum_{k=1}^N \omega_k^{DD/ND}\frac{\phi^{DD/ND}_k(x)\phi^{DD/ND}_k(x')}{2},
\end{split}
\ee
with
\begin{align}
\phi^{DD}_{k}(x) &= \sqrt{\frac{2}{N+1}}\sin (\frac{\pi k\, x}{N+1}), \quad \omega^{DD}_k =2 \sin\left(\frac{\pi k}{2N+2}\right),\\
\phi^{ND}_{k}(x) &= \sqrt{\frac{2}{N+1/2}}\cos (\frac{\pi (k-1/2)}{N+1/2}(x-1/2)), \quad \omega^{ND}_k  = 2\sin\left(\frac{\pi (k-1/2)}{2N+1}\right).
\end{align}
From them, one can reconstruct the correlation functions of the fields as
\be
\begin{split}\label{eq:xil}
X'_{jj'}(x,x') \equiv &\la \Phi^\dagger_j(x)\Phi_{j'}(x')\ra = \l \frac{1+S}{2} \r_{jj'}X_{ND}(x,x') + \l \frac{1-S}{2} \r_{jj'}X_{DD}(x,x'),\\
P'_{jj'}(x,x') \equiv &\la \Pi^\dagger_j(x)\Pi_{j'}(x')\ra = \l \frac{1+S}{2} \r_{jj'}P_{ND}(x,x') + \l \frac{1-S}{2} \r_{jj'}P_{DD}(x,x'),
\end{split}
\ee
regarded as $2N\times 2N$ square matrices. Here, we put an apex on the matrices $X',P'$, referring to the explicit presence of the defect, while we omit it in its absence ($\mathcal{T}=1$).

We focus now on the entanglement properties between the two wires. To do so, we consider the subsystem $A$ made by the first wire, and we construct its reduced correlation matrices, $X'_A$ and $ P'_A$ of dimension $N\times N$ by restricting $j,j' =1$ in Eq.~\eqref{eq:xil}. This allows to write the charged moments of $\rho^n_A$ as
\be\label{eq:ch_mom_bos}
\begin{split}
Z'_n(\alpha) = \text{det}^{-1}\l  (\sqrt{X'_AP'_A}+1/2)^n-e^{i\alpha}(\sqrt{X'_AP'_A}-1/2)^n\r\times \\
\text{det}^{-1}\l  (\sqrt{X'_AP'_A}+1/2)^n-e^{-i\alpha}(\sqrt{X'_AP'_A}-1/2)^n\r.
\end{split}
\ee
Similarly, for $\mathcal{T}=1$ we write it as
\be\label{eq:ch_mom_bosT1}
\begin{split}
Z_n(\alpha) = \text{det}^{-1}\l  (\sqrt{X_AP_A}+1/2)^n-e^{i\alpha}(\sqrt{X_AP_A}-1/2)^n\r\times \\
\text{det}^{-1}\l  (\sqrt{X_AP_A}+1/2)^n-e^{-i\alpha}(\sqrt{X_AP_A}-1/2)^n\r.
\end{split}
\ee
As we did in the fermionic case, the starting point of our analysis is the FCS of the system in the absence of the defect, that is
\cite{mgc-20}
\be\label{eq:charged}
-\log Z_1(\alpha) = \l \frac{|\alpha|}{2\pi}-\l  \frac{\alpha}{2\pi}\r^2 \r  \log \frac{L}{\varepsilon} + \dots, \quad \alpha \in [-\pi,\pi]
\ee
in the large $L/\varepsilon$ limit, and it is periodic under $\alpha \rightarrow \alpha+2\pi$.

\subsection{Spectral density}
In order to evaluate the spectral density, we use a strategy similar to the one employed in Sec.~\ref{sec:spectral_fermi}, using the exact relation between the correlation matrices with and without the defects.

\begin{figure}[t]
\centering
	\includegraphics[width=0.6\linewidth]{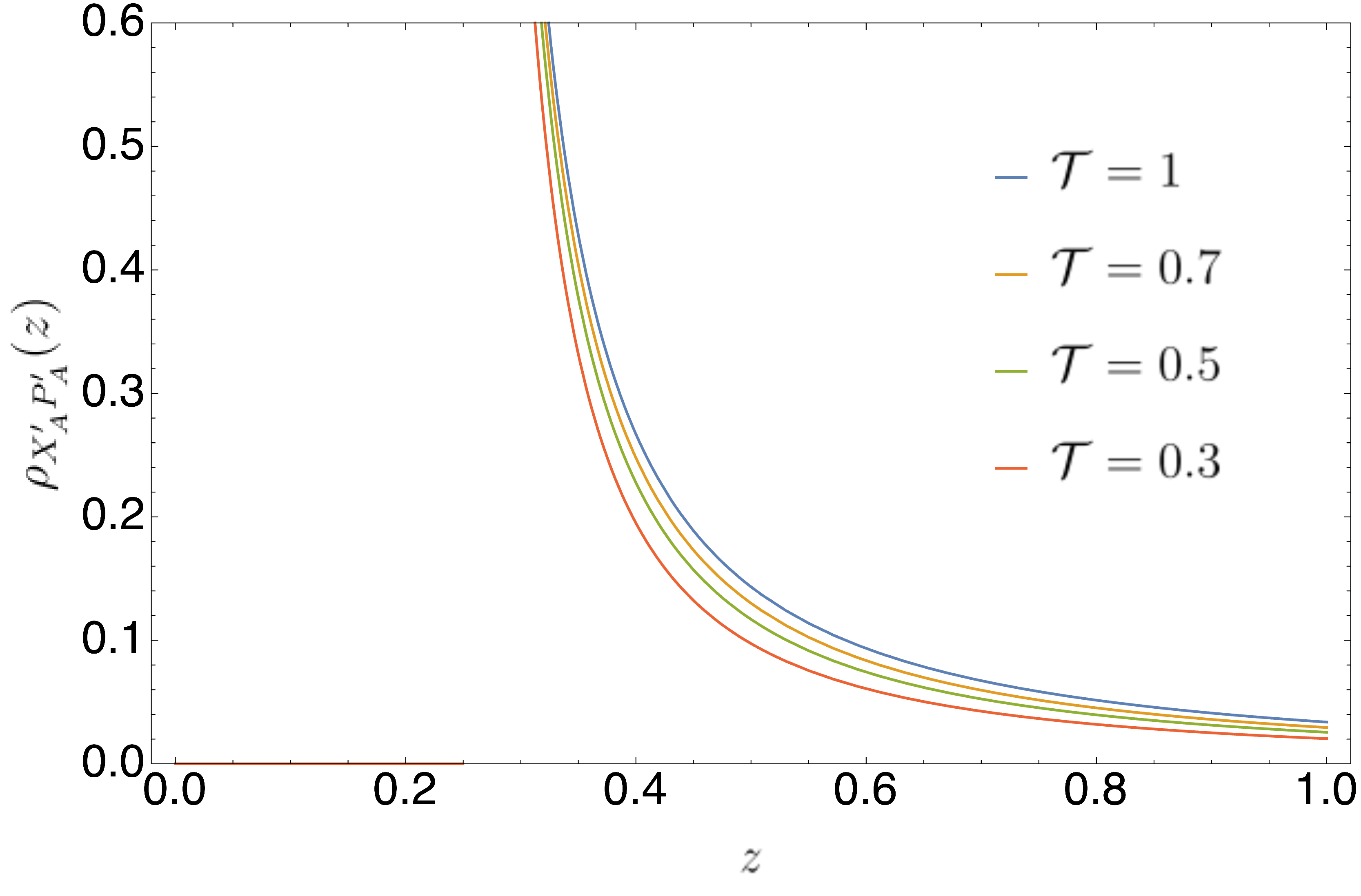}
    \caption{Plot of the analytical prediction of the spectral density of $X'_AP'_A$, given by Eq.~\eqref{eq:SpecDens_Bos} for different values of the transmission probability $\mathcal{T}$. The support of the density is $z \in (1/4,\infty)$, while it vanishes elsewhere.}
    \label{fig:RhoXP_Bos}
\end{figure}

We first define the matrix
\be
E_A = X_AP_A-1/4,
\ee
constructed in the absence of defect, and we use an additional apex when the defect is present, similarly to Eq.~\eqref{eq:EACA}. We repeat the same steps done in Sec.~\ref{sec:spectral_fermi}: we derive the result for $E_A$, then for $E_A'$ and by inverting the relation between $E_A'$ and $X'_AP'_A$, we recover the desired result.
For $n=1$, one can easily show that Eqs.~\eqref{eq:ch_mom_bosT1},~\eqref{eq:ch_mom_bos} can be rewritten as
\be\label{eq:FCS_boson}
\begin{split}
Z_1(\alpha) = \text{det}^{-1}\l  1+4\sin^2(\alpha/2)(X_AP_A-1/4)\r,\\
Z'_1(\alpha) = \text{det}^{-1}\l  1+4\sin^2(\alpha/2)(X'_AP'_A-1/4)\r.
\end{split}
\ee 
From the above equations, it is possible to write a relation between the resolvent of $E_A$, denoted by $G_{E_A}$, and the full counting statistics $Z_1(\alpha)$ as
\be\label{eq:s_boson}
G_{E_A}(y) = -\frac{d}{dy} \log Z_1(\alpha(y)),
\ee
where $y = -1/(4\sin^2 (\alpha/2))$.
We notice that the relation above slightly differs from Eq.~\eqref{eq:Gfermions}, valid for fermions, due to the presence of a different multiplicative factor and a minus sign in the definition of $y$. Inserting Eq.~\eqref{eq:charged} in Eq.~\eqref{eq:s_boson}, one gets after straightforward algebra
\be
G_{E_A}(y) = \frac{1}{4\pi y^2 \sqrt{1+\frac{1}{4y}}\sqrt{-\frac{1}{y}}} - \frac{\text{arcsin}\sqrt{-\frac{1}{4y}}}{2\pi^2 y^2 \sqrt{1+\frac{1}{4y}} \sqrt{-\frac{1}{y}}}.
\ee
As expected, the resolvent has a branch-cut for $y \in (0,+\infty)$, corresponding to the support of the spectrum of $E_A$, and it is smooth elsewhere. In particular, for those values of $y$ we get
\be
\rho_{E_A}(y)  \equiv \frac{1}{\log(L/\varepsilon)}\frac{1}{2\pi}\text{Im}\l G_{E_A}(y-i0^+) - G_{E_A}(y+i0^+)\r = \frac{1}{2\pi^2 y \sqrt{1+4y}}.
\ee
A remarkable relation, which follows directly from the definition (see also~\cite{cmc-22a}), is 
\be
E_A' = \mathcal{T}E_A,
\ee
which allows us to recover the spectral density with the defect via a simple rescaling, as
\be
G_{E'_A}(y) = G_{E_A}(y/\mathcal{T}), \quad \rho_{E'_A}(y) = \frac{1}{\mathcal{T}}\rho_{E_A}(y/\mathcal{T}).
\ee
For the sake of completeness, we express the spectral density of the matrix $X'_A P'_A$, related to $E'_A$ as
\be
X'_A P'_A=E'_A+1/4,
\ee
and we finally obtain 
\be\label{eq:SpecDens_Bos}
\rho_{X'_A P'_A}(z) = \begin{cases}\frac{1}{2\pi^2}\frac{1}{(z-1/4)\sqrt{1+(4z-1)/\mathcal{T}}}, \quad z \in (1/4,\infty),\\
0, \quad \text{otherwise},
\end{cases}
\ee
that we plot in Fig.~\ref{fig:RhoXP_Bos} for different values of $\mathcal{T}$.
A striking application of Eq.~\eqref{eq:SpecDens_Bos} is the computation of the charged moments for any value of $n$ (even non-integer), which read
\be\label{eq:FCSnB}
-\frac{\log Z'_n(\alpha)}{\log(L/\varepsilon)} = \int^\infty_{1/4}
 dz \rho_{X'_A P'_A}(z)\log((\sqrt{z}-1/2)^{2n}+(\sqrt{z}+1/2)^{2n}-2\cos \alpha(z-1/4)^{n}).
 \ee
 As a byproduct, we compute the FCS in the presence of the defect as
\begin{multline}
\label{eq:z1nb}
-\log Z'_1(\alpha) = \l \frac{|\text{arcsin}(\sqrt{\mathcal{T}} \sin(\alpha/2))|}{\pi}-\l  \frac{\text{arcsin}(\sqrt{\mathcal{T}} \sin(\alpha/2))}{\pi}\r^2 \r  \log \frac{L}{\varepsilon} + O(1), \\ \quad \alpha \in [-\pi,\pi],
\end{multline}
a result which was already obtained with completely different methods in Ref.~\cite{cmc-22a}. Indeed, as we did in Eq.~\eqref{eq:prodfermions} for fermions, we can write the charged moments in the factorized form
\be
Z'_n(\alpha) = \prod^{n-1}_{p=0} Z'_1(2\pi p/n + \alpha/n),
\ee
and using the result in Eq.~\eqref{eq:z1nb}, we get 
\begin{multline}
-\log Z'_n(\alpha) = \log(L/\varepsilon)\times\\ \sum^{n-1}_{p=0}\l \left|\frac{\text{arcsin}(\sqrt{\mathcal{T}} \sin(\pi p/n + \alpha/2n))}{\pi}\right|-\l  \frac{\text{arcsin}(\sqrt{\mathcal{T}} \sin(\pi p/n +\alpha/2n))}{\pi}\r^2 \r.
\end{multline}
We finally notice that the last expression, valid for any $\alpha$, has a singularity at $\alpha=0$ for any value of $\mathcal{T} \in (0,1]$, which comes from the term at $p=0$.

\begin{figure}[t]
\centering
	{\includegraphics[width=0.49\linewidth]{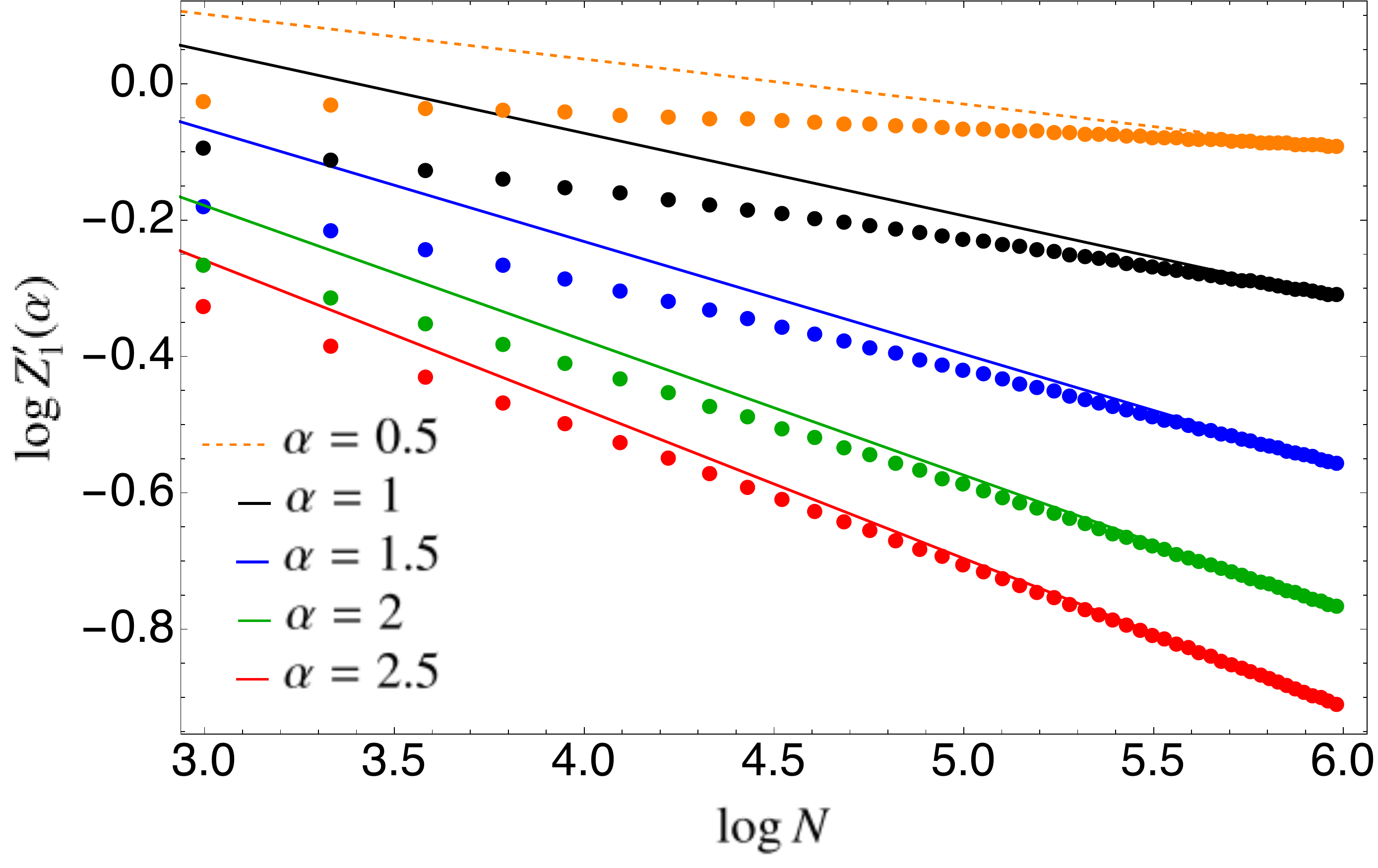}}
 \subfigure
 {\includegraphics[width=0.49\linewidth]{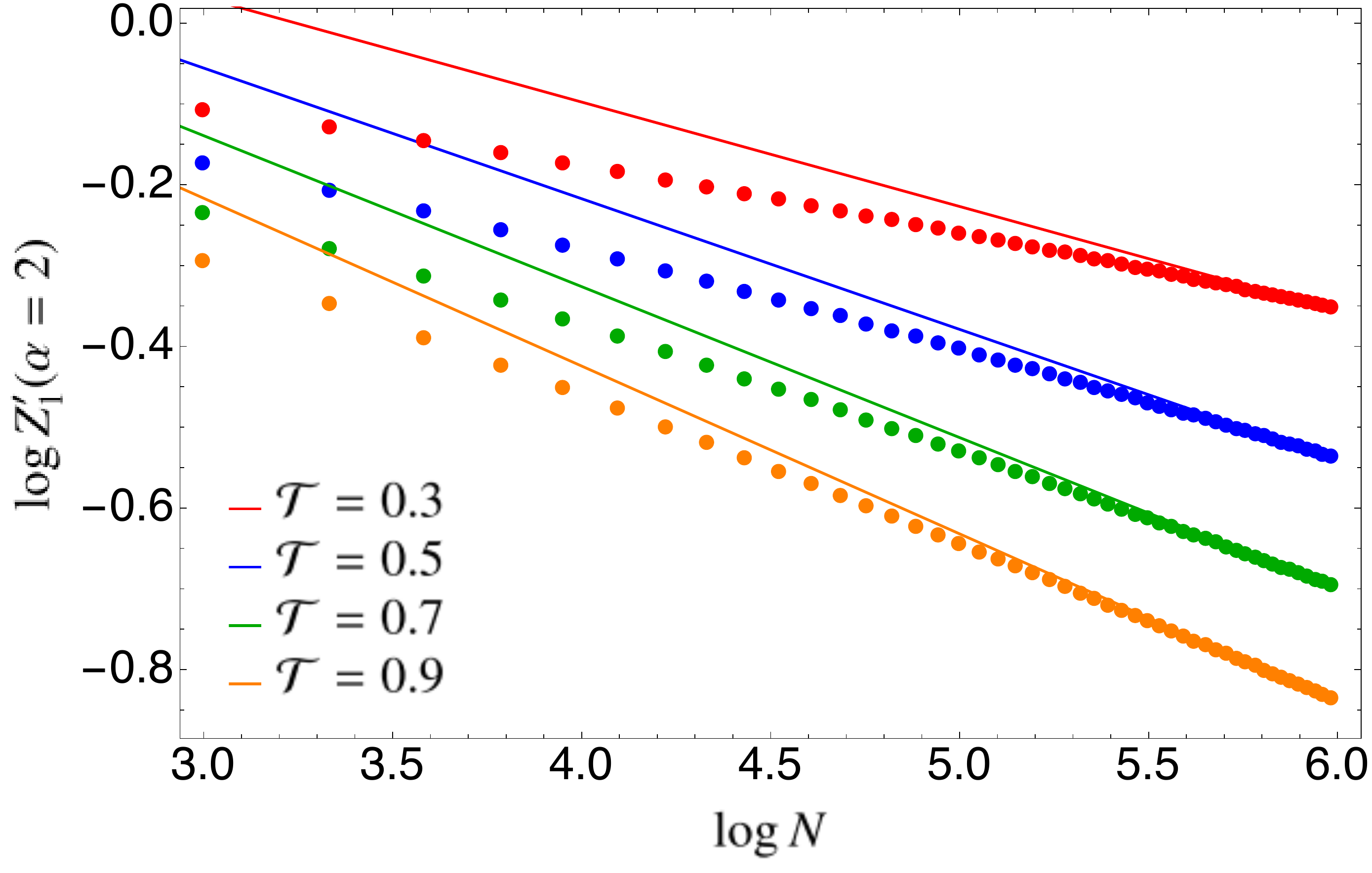}}
    \caption{Full counting statistics as a function of the (logarithmic) subsystem size, $N$. The left panel shows the behaviour for different values of $\alpha$ at fixed $\mathcal{T}=0.8$, while the right panel have been obtained fixing $\alpha=2$ and varying $\mathcal{T}$. The solid lines correspond to Eq. \eqref{eq:FCSnB}, where we have fitted the constant term. The analytical formula matches well the data as $N$ increases and for large values of $\alpha$ and $\mathcal{T}$, but for values of the parameters much larger values of $N$ are necessary to observe a fair match, as explained in the text.
    }
    \label{fig:n2}
\end{figure}

In Fig.~\ref{fig:n2} we test the prediction for the full counting statistics given by Eq.~\eqref{eq:FCSnB}. We find that the agreement improves as the subsystem size $N$ increases and for large values of $\alpha$ and $\mathcal{T}$. It was already proven in~\cite{mgc-gapped,mgc-20} that, without a defect, the lattice results for the charged moments of the complex harmonic chains approach the theoretical ones in a nonuniform way. Indeed, the subleading contributions in $L$ as function of $\alpha$ are not known for bosonic theories and this prevents us from finding always a good match with the numerics (see, e.g., the dashed orange line in the left panel of Fig.~\ref{fig:n2}). This scenario further worsens as $\mathcal{T}$ approaches zero and the two chains become decoupled: in this case, the coefficient of the logarithmic term goes to $0$, but for small and finite values of $\mathcal{T}$ the corrections become of the same order as the leading order term, therefore larger values of the system size are required to see a good agreement (see the red line in the right panel of Fig.~\ref{fig:n2}).


\subsection{Symmetry resolved entropies}

As an application of the results for the charged moments, we can find the symmetry resolved R\'enyi entropies in the scaling limit $L/\varepsilon \gg 1$. To do so, we need to understand the behaviour of the charged moments near $\alpha=0$. Here the situation is more subtle with respect to the fermionic case, as a cusp singularity is present at $\alpha=0$, and one cannot apply the saddle point approximation. Indeed, an expansion for small values of $\alpha$ is given by 
\be\label{eq:chargedb}
-\log Z'_n(\alpha) = (A_n(\mathcal{T}) + B_n(\mathcal{T})|\alpha| + O(\alpha^2))\log \frac{L}{\varepsilon} +\dots,
\ee
with $A_n(\mathcal{T}),B_n(\mathcal{T})$ given by, respectively,
\be
A_n(\mathcal{T}) =  \int^\infty_{1/4}
 dz \rho_{X'_A P'_A}(z)\log((\sqrt{z}-1/2)^{2n}+(\sqrt{z}+1/2)^{2n}-2(z-1/4)^{n}),
\ee
(again for integer $n$ a simpler expression can be found expanding Eq. \eqref{eq:z1nb}) and
\be
B_n(\mathcal{T}) = \frac{\sqrt{\mathcal{T}}}{2\pi n}.
\ee
Once we plug the result of Eq.~\eqref{eq:chargedb} in the Fourier transform~\eqref{eq:FT}, we obtain
\begin{align}\label{eq:symmB}
{\cal Z}_n(q) =& Z_n(0)\frac{B_n(\mathcal{T})\log L/\varepsilon}{\pi(q^2+(B_n(\mathcal{T})\log L/\varepsilon)^2)},\\
S_n(q) =& S_n -\log \log \frac{L}{\varepsilon} + O(1).
\end{align}
 It is worth to mention that we obtain the equipartition of the entanglement at leading order, with a double logarithmic correction with a prefactor $1$ rather than $1/2$, as it happens for the fermions in Eq.~\eqref{eq:symmF}. On the other hand, the first term breaking the equipartition is $O(q^2/(\log \frac{L}{\varepsilon})^2)$, which differs from the $O(q^2/\log N)$ found for fermions in Eq.~\eqref{eq:symmF}~\cite{mgc-20,hcc-21}.

\section{Conclusions}\label{sec:conclusions}

In this work, we have investigated the quantum fluctuations of the $U(1)$ charge for free massless boson and fermions in the presence of an interface defect. In particular, we have considered a family of conformal defects, parametrised by the transmission probability $\mathcal{T}$, giving analytical predictions for the full-counting statistics, the charged moments and the symmetry resolved R\'enyi entropies. Our approach relies on finding the spectral densities of the correlation matrix of the models restricted to the subsystem, by relating what happens when $\mathcal{T}<1$ to the case $\mathcal{T}=1$, i.e. in the absence of the defect.
This technique is complementary to the standard replica trick~\cite{cc-04} and its main advantage is that it provides automatically an analytical continuation for any non-integer R\'enyi index $n$.
We also emphasize that, while some universal features depend explicitly on $\mathcal{T}$, as the logarithmic prefactor of $\log Z'_n(\alpha)$, the leading terms of the symmetry resolved entropies depends on ${\cal T}$ only through the total entropy (except for the trivial case $\mathcal{T}=0$ where the wires are decoupled). 

We mention that our result can be trivially adapted to tackle junctions with multiple wires. In particular, as shown in Refs.~\cite{cmc-22,cmc-22a} using conformal field theory tools, it is possible to relate directly any non-trivial bipartition of wires to the simpler case of two wires. Moreover, still from the results of Refs.~\cite{cmc-22,cmc-22a}, one can also derive the symmetry resolved negativity~\cite{goldstein1,mbc-21} between two wires in a three-wire geometry.

An interesting problem that we leave for future investigation is the possible relation between the spectral properties of the reduced correlation matrix (studied in this paper) and the ones of the reduced density matrix $\rho_A$, an important quantity known as entanglement spectrum~\cite{spectrum}. A close direction is the characterisation of the modular flow in the presence of a defect, which has been partially addressed in~\cite{mt2-21}. 

\section*{Acknowledgements}
PC and LC acknowledge support from ERC under Consolidator grant number 771536 (NEMO). SM thanks support from Institute for Quantum Information and Matter and the Walter Burke Institute for Theoretical Physics at Caltech.

\appendix

\section{Finite-dimensional representation of the correlation function}\label{app:proof1}

In this appendix, we explain how to represent the correlation function in \eqref{eq_Kernel} as a finite dimensional matrix. The method has been considered in Ref. \cite{cmc-22}, and we report here the details of the derivation.\\ From now on, we set the spatial size to be $L=1$. Thus, the kernel $C'_{ij}(x,y)$ can be regarded as an operator acting on the Hilbert space $\mathbb{C}^2 \otimes L^2([0,1])$. In particular, the factor $L^2([0,1])$ represents the wave-functions on $[0,1]$, while $\mathbb{C}^2$ labels the space of the two wires. The first issue is that the Hilbert space $L^2([0,1])$ is infinite dimensional, and the operator $C'_{ij}(x,y)$ has an infinite number of eigenvalues. However, both $C_{ND}$ and $C_{DD}$ appearing in Eq. \eqref{eq_Kernel} act non-trivially only on a finite-dimension subspace $\mathcal{H}_0$ (their image is finite dimensional). The latter property is important, since only the non-zero eigenvalues of the kernel $C'$ enter the evaluation of the entanglement measures we are interested in.

In particular, we choose $\mathcal{H}_0     \subset L^2([0,1])$ to be generated by the independent wavefunctions
\be
e_n \equiv \begin{cases} \phi^{DD}(n,x), \quad 1\leq n \leq N \\ \phi^{ND}(n-N,x), \quad 1+N\leq n \leq 2N,
\label{eq_en}
\end{cases}
\ee
that we pick as a (non-orthonormal) basis. It is easy to show that, since the single-particle eigenfunctions are normalised in both ND and DD sectors, one has
\be
\la e_n,e_{n'} \ra = \int dx \  \overline{\phi^{DD}(n,x)}\phi^{DD}(n',x) = \delta_{n,n'}, \quad n,n' = 1,\dots,N,
\ee
and similarly if $n,n' = 1+N,\dots,2N$. However,
 for $n\leq N$ and $n'\geq N+1$, their scalar product $Q_{n,n'}$ is non-vanishing, and it is
\be
Q_{n,n'} \equiv \la e_n,e_{n'} \ra  = 2 \int^1_0 dx \sin\l n\pi x\r\cos\l \l n' -\frac{1}{2}\r\pi x\r = \frac{2 n}{\pi\l n^2-(n'-1/2)^2\r}.
\label{eq_MatrixQ1}
\ee
We now aim to give a matrix representation of the kernels $C_{ND},C_{DD}$ in the subspace $\mathcal{H}_0$, and one needs to be careful since the basis in not orthonormal. To overcome this issue, we introduce the dual space $\mathcal{H}^*_0$, namely the space of linear functional on $\mathcal{H}_0$, and the dual basis associated to $\{e_n\}$
\be
\{e^*_n\}_{n=1,\dots,2N},
\ee
that is
\be
e^*_n(e_{n'}) = \delta_{nn'},
\ee
by definition. To avoid confusion, we use another symbol for the bra associated to $e_n$, denoted by
\be
e^\dagger_n \in \mathcal{H}^*_0,
\ee
and satisfying
\be
e^\dagger_n(v) \equiv \la e_n,v \ra, \quad \forall v \in \mathcal{H}_0.
\ee
Both $e^\dagger_n$ and $e^*_n$ belong to $\mathcal{H}_0$, but they are different since
\be
e^\dagger_n(e_{n'}) = Q_{n,n'}, \quad  e^*_n(e_{n'}) = \delta_{nn'}  ,\quad n\leq N, n'\geq N+1,
\ee
which is a consequence of the non-orthonormality of the basis.

We now express the projectors $C_{ND}$ and $C_{DD}$, seen as operators of $\mathcal{H}_0$ and belonging to
\be
\text{End}\l \mathcal{H}_0 \r \simeq \mathcal{H}_0\otimes  \mathcal{H}^*_0,
\ee
in terms of the basis as
\be
C_{DD} = \sum^{N}_{n=1} e_n\otimes e^\dagger_n, \quad C_{ND} = \sum^{2N}_{n=N+1} e_n\otimes e^\dagger_n.
\ee
In particular, their basis elements can be computed as
\be
\begin{split}
&\l C_{DD} \r_{n,n'} \equiv e^*_n(C_{DD} \ e_{n'}) = \sum^N_{m=1} e^*_n(e_m) e_m^\dagger(e_{n'}) = \sum^N_{m=1} \delta_{n,m} \la e_m,e_{n'} \ra\\
&\l C_{ND} \r_{n,n'} \equiv e^*_n(C_{ND} \ e_{n'}) = \sum^{2N}_{m=N+1} e^*_n(e_m) e_m^\dagger(e_{n'}) = \sum^{2N}_{m=N+1} \delta_{n,m} \la e_m,e_{n'} \ra,
\end{split}
\ee
and their $2N\times 2N$ matrix takes the following expression
\be
C_{DD} = \begin{pmatrix} 1 & Q\\ 0 & 0\end{pmatrix}, \quad C_{ND}= \begin{pmatrix} 0 & 0\\ Q^\dagger & 1\end{pmatrix}.
\label{CDDN}
\ee
Putting everything together, we are able to express the correlation function $C'_{ij}(x,y)$ in Eq. \eqref{eq_Kernel} as a $4N\times 4N$ matrix acting on 
\be
\mathbb{C}^{2}\otimes \mathcal{H}_0 \simeq \mathbb{C}^{2}\otimes \mathbb{C}^{2N},
\ee
as
\be
C' =\frac{1-S}{2} \otimes \begin{pmatrix} 1 & Q\\ 0 & 0\end{pmatrix} + \frac{1+S}{2} \otimes \begin{pmatrix} 0 & 0\\ Q^\dagger & 1\end{pmatrix}.
\ee
that is the result we employed in Eq. \eqref{eq_MatrixC} of the main text.

\subsection{Proof of Eq. \eqref{eq:trEA}}\label{app:proof2}

In this Appendix we  prove Eq. \eqref{eq:trEA} following Ref. \cite{cmc-22}. As a first step, we introduce an auxiliary quantity for the computation of the entanglement entropy, the covariance matrix $\Gamma' = 1-2C'$ of dimension $4N\times 4N$. 
We restrict $\Gamma'$ to the subsystem $A$ made of one wire, and we get the $2N\times 2N$ matrix $\Gamma'_A = 1-2C'_A$. Employing the finite-dimensional representation of the correlation matrix  $C'_A$ in Eq. \eqref{eq:CA} toghether with \eqref{CDDN}, and making use of the property $(C_{DD})^2=C_{DD}, (C_{ND})^2=C_{ND}$ ($C_{DD}$ and $C_{ND}$ are projectors), we have \be\label{eq:EA_T}
1-{\Gamma'_{A}}^2 =4E'_A= \mathcal{T} (C_{ND}-C_{DD})^2,
\ee
where $E'_A$ is defined by \eqref{eq:EACA} and $\mathcal{T}$ is the transmission probability.
In particular, for the completely transmitting $S$-matrix obtained with $\mathcal{T}=1$
\be
S = \begin{pmatrix} 0 & 1 \\ 1 & 0
\end{pmatrix},
\ee
we get
\be\label{eq:EA_T1}
4E_A=(C_{ND}-C_{DD})^2.
\ee
Thus, for a permeable defect with $\mathcal{T}<1$, one finds that the matrix $E'_A$ is related to the one appearing in Eq. \eqref{eq:EA_T1} via the rescaling factor $\mathcal{T}$, as shown by Eq. \eqref{eq:EA_T}, that is the result in Eq.
\eqref{eq:trEA} of the main text.


\begin{thebibliography}{}

 \bibitem{ll-93}
L.S. Levitov and G.B. Lesovik, \textit{Charge distribution in quantum shot
noise}, \href{https://ui.adsabs.harvard.edu/abs/1993JETPL..58..230L/abstract}{JETP Lett. {\bf 58}, 230 (1993).}

\bibitem{s-07}
K. Schonhamer, \textit{Full counting statistics for non-interacting fermions: exact results and the Levitov-Lesovik formula}, \href{
https://doi.org/10.1103/PhysRevB.75.205329}{Phys. Rev. B \textbf{75}, 205329
(2007).}

\bibitem{cd-07}
R. W. Cherng and E. Demler, {\it Quantum Noise Analysis of Spin Systems Realized with Cold Atoms},
\href{http://dx.doi.org/10.1088/1367-2630/9/1/007}{New J. Phys. {\bf 9}, 7 (2007)}.

\bibitem{bss-07}
M. Bortz, J. Sato, and M. Shiroishi M, {\it String correlation functions of the spin-1/2 Heisenberg XXZ chain},  
\href{https://doi.org/10.1088/1751-8113/40/16/001}{J. Phys. A {\bf 40}, 4253 (2007)}.

\bibitem{aem-08}
D. B. Abraham, F. H. L. Essler, and A. Maciolek, {\it Effective Forces Induced by a Fluctuating Interface: Exact Results}, 
\href{https://doi.org/10.1103/PhysRevLett.98.170602}{Phys. Rev. Lett. {\bf 98}, 170602 (2007)}.

\bibitem{lp-08}
A. Lamacraft and P. Fendley, {\it Order Parameter Statistics in the Critical Quantum Ising Chain},
\href{https://doi.org/10.1103/PhysRevLett.100.165706}{Phys. Rev. Lett. {\bf 100}, 165706 (2008)}.

\bibitem{ehm-09}
M. Esposito, U. Harbola and S. Mukamel, \textit{Nonequilibrium fluctuations,
fluctuation theorems, and counting statistics in quantum systems}, \href{
https://doi.org/10.1103/RevModPhys.81.1665}{Rev.
Mod. Phys. \textbf{81}, 1665-1702 (2009).}


\bibitem{ia-13}
D. A. Ivanov and A. G. Abanov, {\it Characterizing correlations with full counting statistics: Classical Ising and quantum XY spin chains},
\href{http://dx.doi.org/10.1103/PhysRevE.87.022114}{Phys. Rev. E {\bf 87}, 022114 (2013)}.

\bibitem{sk-13}
Y. Shi and I. Klich, {\it Full counting statistics and the Edgeworth series for matrix product states},
\href{http://dx.doi.org/10.1088/1742-5468/2013/05/P05001}{J. Stat. Mech. (2013) P05001.}

\bibitem{e-13}
V. Eisler, {\it Universality in the Full Counting Statistics of Trapped Fermions},
\href{https://doi.org/10.1103/PhysRevLett.111.080402}{Phys. Rev. Lett. {\bf 111}, 080402 (2013).}

\bibitem{k-14}
I. Klich, {\it A note on the Full Counting Statistics of paired fermions},
\href{http://dx.doi.org/10.1088/1742-5468/2014/11/P11006}{J. Stat. Mech. (2014) P11006}.

\bibitem{mcsc-15}
M. Moreno-Cardoner, J. F. Sherson and G. De Chiara, 
{\it Non-Gaussian distribution of collective operators in quantum spin chains}, 
\href{http://dx.doi.org/10.1088/1367-2630/18/10/103015}{New J. Phys. {\bf 18}, 103015 (2016).}

\bibitem{bd-16}
D. Bernard and B. Doyon, \textit{Conformal Field Theory out of Equilibrium: a
Review}, \href{https://iopscience.iop.org/article/10.1088/1742-5468/2016/06/064005}{J. Stat. Mech. 064005 (2016).}


\bibitem{sp-17}
J.-M. St\'ephan and F. Pollmann, {\it Full counting statistics in the Haldane-Shastry chain},
\href{http://dx.doi.org/10.1103/PhysRevB.95.035119}{Phys. Rev. B {\bf 95}, 035119 (2017).}
	
\bibitem{CoEG17} 
M. Collura, F. H. L. Essler, and S. Groha, {\it Full counting statistics in the spin-1/2 Heisenberg XXZ chain},
\href{http://dx.doi.org/10.1088/1751-8121/aa87dd}{J. Phys. A {\bf 50}, 414002 (2017).}

\bibitem{nr-17}
K. Najafi and M. A. Rajabpour, {\it Full counting statistics of the subsystem energy for free fermions and quantum spin chains},
\href{http://dx.doi.org/10.1103/PhysRevB.96.235109}{Phys. Rev. B {\bf 96}, 235109 (2017).}

\bibitem{hb-17}
S. Humeniuk and H. P. B\"uchler, {\it Full Counting Statistics for Interacting Fermions with Determinantal Quantum Monte Carlo Simulations},
\href{http://dx.doi.org/10.1103/PhysRevLett.119.236401}{Phys. Rev. Lett. {\bf 119}, 236401 (2017)}.

\bibitem{bpc-18}
A. Bastianello, L. Piroli, and P. Calabrese, 
{\it Exact local correlations and full counting statistics for arbitrary states of the one-dimensional interacting Bose gas},
\href{http://dx.doi.org/10.1103/PhysRevLett.120.190601}{Phys. Rev. Lett. {\bf 120}, 190601 (2018).}

\bibitem{bp-18}
A. Bastianello and L. Piroli, {\it  From the sinh-Gordon field theory to the one-dimensional Bose gas: exact local correlations and full counting statistics},
\href{http://dx.doi.org/10.1088/1742-5468/aaeb48}{ J. Stat. Mech. (2018) 113104}.

\bibitem{ppg-19}
G. Perfetto, L. Piroli, and Andrea Gambassi,
{\it Quench action and large deviations: Work statistics in the one-dimensional Bose gas},
\href{https://doi.org/10.1103/PhysRevE.100.032114}{Phys. Rev. E {\bf 100}, 032114 (2019)}. 

\bibitem{dbdm-20}
G. Del Vecchio Del Vecchio, A. Bastianello, A. De Luca, G. Mussardo,
{\it Exact out-of-equilibrium steady states in the semiclassical limit of the interacting Bose gas},
\href{http://dx.doi.org/10.21468/SciPostPhys.9.1.002}{SciPost Phys. {\bf 9}, 002 (2020)}. 

\bibitem{dm-20}
B. Doyon and J. Myers, \textit{Fluctuations in ballistic transport from Euler
hydrodynamics}, \href{
https://doi.org/10.1007/s00023-019-00860-w
}{Ann. Henri Poincare \textbf{21}, 255 (2020).}


\bibitem{gadp-06}
V. Gritsev, E. Altman, E. Demler and A. Polkovnikov, 
{\it Full quantum distribution of contrast in interference experiments between interacting one-dimensional Bose liquids},
\href{http://dx.doi.org/10.1038/nphys410}{Nature Phys. {\bf 2}, 705 (2006)}.



\bibitem{er-13}
V. Eisler and Z. Racz, {\it Full Counting Statistics in a Propagating Quantum Front and Random Matrix Spectra},
\href{https://doi.org/10.1103/PhysRevLett.110.060602}{Phys. Rev. Lett. {\bf 110}, 060602 (2013)}.


\bibitem{lddz-15}
I. Lovas, B. Dora, E. Demler, and G. Zarand, {\it Full counting statistics of time of flight images},
\href{http://dx.doi.org/10.1103/PhysRevA.95.053621}{Phys. Rev. A {\bf 95}, 053621 (2017)}.

\bibitem{gec-18}
S. Groha, F. H. L. Essler, and P. Calabrese, {\it Full counting statistics in the transverse field Ising chain},
\href{https://doi.org/10.21468/SciPostPhys.4.6.043}{SciPost Phys. {\bf 4}, 043 (2018)}.

\bibitem{ce-20}
M. Collura and F. H. L. Essler, {\it How order melts after quantum quenches}, 
\href{http://dx.doi.org/10.1103/PhysRevB.101.041110}{Phys. Rev. B {\bf 101}, 041110 (2020)}.

\bibitem{c-19}
M. Collura, {\it Relaxation of the order-parameter statistics in the Ising quantum chain},
\href{http://dx.doi.org/10.21468/SciPostPhys.7.6.072}{SciPost Phys. {\bf 7}, 072 (2019)}.

\bibitem{ag-19}
M. Arzamasovs and D. M. Gangardt, {\it Full Counting Statistics and Large Deviations in a Thermal 1D Bose Gas},
\href{https://doi.org/10.1103/PhysRevLett.122.120401}{Phys. Rev. Lett. {\bf 122}, 120401 (2019)}.

\bibitem{nr-20}
M. N. Najafi and M. A. Rajabpour, {\it Formation probabilities and statistics of observables as defect problems in the free fermions and the quantum spin chains},
\href{https://doi.org/10.1103/PhysRevB.101.165415}{Phys. Rev. B {\bf 101}, 165415 (2020)}.



\bibitem{cgcm-20}
P. Calabrese, M. Collura, G. Di Giulio and S. Murciano, 
{\it Full counting statistics in the gapped XXZ spin chain},
\href{https://iopscience.iop.org/article/10.1209/0295-5075/129/60007}{EPL {\bf 129},  60007 (2020)}.

\bibitem{vcc-20}
R. J. Valencia Tortora, P. Calabrese, and M. Collura, {\it Relaxation of the order-parameter statistics and dynamical confinement},
\href{https://doi.org/10.1209/0295-5075/132/50001}{EPL {\bf 132}, 50001 (2020)}. 

\bibitem{cdcd-20}
M. Collura, A. De Luca, P. Calabrese, and J. Dubail, {\it Domain-wall melting in the spin-1/2 XXZ spin chain: emergent Luttinger liquid with fractal quasi-particle charge},
\href{https://doi.org/10.1103/PhysRevB.102.180409}{Phys. Rev. B {\bf 102}, 180409(R) 2020}.





\bibitem{bcckr-22}
B. Bertini, P. Calabrese, M. Collura, K. Klobas, and C. Rylands,
{\it Nonequilibrium Full Counting Statistics and Symmetry-Resolved Entanglement from Space-Time Duality},
\href{https://arxiv.org/abs/2212.06188}{arXiv:2212.06188}.



\bibitem{AJKB10} 
J. Armijo, T. Jacqmin, K. V. Kheruntsyan, and I. Bouchoule, 
{\it Probing Three-Body Correlations in a Quantum Gas Using the Measurement of the Third Moment of Density Fluctuations},
\href{http://dx.doi.org/10.1103/PhysRevLett.105.230402}{Phys. Rev. Lett. {\bf 105}, 230402 (2010)}.

\bibitem{JABK11} 
T. Jacqmin, J. Armijo, T. Berrada, K. V. Kheruntsyan, and I. Bouchoule, 
{\it Sub-Poissonian Fluctuations in a 1D Bose Gas: From the Quantum Quasicondensate to the Strongly Interacting Regime},
\href{http://dx.doi.org/10.1103/PhysRevLett.106.230405}{Phys. Rev. Lett. {\bf 106}, 230405 (2011)}.

\bibitem{HLSI08} 
S. Hofferberth, I. Lesanovsky, T. Schumm, A. Imambekov, V. Gritsev, E. Demler, and J. Schmiedmayer, 
{\it Probing quantum and thermal noise in an interacting many-body system},
\href{http://dx.doi.org/10.1038/nphys941}{Nature Phys. {\bf 4}, 489 (2008)}.

\bibitem{KPIS10} 
T. Kitagawa, S. Pielawa, A. Imambekov, J. Schmiedmayer, V. Gritsev, and E. Demler, 
{\it Ramsey Interference in One-Dimensional Systems: The Full Distribution Function of Fringe Contrast as a Probe of Many-Body Dynamics},
\href{http://dx.doi.org/10.1103/PhysRevLett.104.255302}{Phys. Rev. Lett. {\bf
    104}, 255302 (2010)}. 

\bibitem{KISD11} 
T. Kitagawa, A. Imambekov, J. Schmiedmayer, and E. Demler, 
{\it The dynamics and prethermalization of one-dimensional quantum systems probed through the full distributions of quantum noise},
\href{http://dx.doi.org/10.1088/1367-2630/13/7/073018}{New J. Phys. {\bf 13}, 73018 (2011).}

\bibitem{GKLK12} 
M. Gring, M. Kuhnert, T. Langen, T. Kitagawa, B. Rauer, M. Schreitl, I. Mazets, D. A. Smith, E. Demler, and J. Schmiedmayer, 
{\it Relaxation and Prethermalization in an Isolated Quantum System},
\href{http://dx.doi.org/10.1126/science.1224953}{Science {\bf 337}, 1318 (2012)}.


\bibitem{zyz-22}
T. Zhou, K. Yang, Z. Zhu, X. Yu, S. Yang, W. Xiong, X. Zhou, X. Chen, C. Li, J. Schmiedmayer, X. Yue, and Y. Zhai, 
{\it Observation of atom-number fluctuations in optical lattices via quantum collapse and revival dynamics}
\href{http://dx.doi.org/10.1103/PhysRevA.99.013602}{ Phys. Rev. A {\bf 99}, 013602 (2019)}.

\bibitem{kl-09}
I. Klich and L. Levitov, {\it Quantum Noise as an Entanglement Meter},
\href{https://doi.org/10.1103/PhysRevLett.102.100502}{Phys. Rev. Lett. {\bf 102}, 100502 (2009).}

\bibitem{kl-09b}
I. Klich and L. Levitov, {\it Many-Body Entanglement: a New Application of the Full Counting Statistics},
\href{https://doi.org/10.1063/1.3149497}{Adv. Theor. Phys. {\bf 1134},
  36 (2009).}

\bibitem{hgf-09}
B. Hsu, E. Grosfeld, and E. Fradkin, {\it Quantum noise and entanglement generated by a local quantum quench},
\href{https://doi.org/10.1103/PhysRevB.80.235412}{Phys. Rev. B {\bf 80}, 235412 (2009)}.

\bibitem{sfr-11a}
H. F. Song, C. Flindt, S. Rachel, I. Klich, and K. Le Hur,
{\it Entanglement from Charge Statistics: Exact Relations for Many-Body Systems},
\href{http://dx.doi.org/10.1103/PhysRevB.83.161408}{Phys. Rev. B {\bf 83}, 161408(R) (2011).}

\bibitem{sfr-11b}
H. F. Song, S. Rachel, C. Flindt, I. Klich, N. Laflorencie, and K. Le Hur,
{\it Bipartite Fluctuations as a Probe of Many-Body Entanglement},
\href{http://dx.doi.org/10.1103/PhysRevB.85.035409}{Phys. Rev. B {\bf 85}, 035409 (2012).}

\bibitem{cmv-12fcs}
P. Calabrese, M. Mintchev and E. Vicari, {\it Exact relations between particle fluctuations and entanglement in Fermi gases},
\href{http://dx.doi.org/10.1209/0295-5075/98/20003}{EPL {\bf 98}, 20003 (2012).}

\bibitem{lbb-12}
G. C. Levine, M. J. Bantegui, and J. A. Burg, {\it Full counting statistics in a disordered free fermion system},
\href{https://doi.org/10.1103/PhysRevB.86.174202}{Phys. Rev. B {\bf 86}, 174202 (2012)}.

\bibitem{si-13}
R. Susstrunk and D. A. Ivanov,
{\it Free fermions on a line: Asymptotics of the entanglement entropy and entanglement spectrum from full counting statistics},
\href{https://doi.org/10.1209/0295-5075/100/60009}{EPL {\bf 100}, 60009 (2012).}

\bibitem{clm-15}
P. Calabrese, P. Le Doussal, and S. N. Majumdar, {\it Random matrices and entanglement entropy of trapped Fermi gases},
\href{https://doi.org/10.1103/PhysRevA.91.012303}{Phys. Rev. A {\bf 91}, 012303 (2015).}

\bibitem{u-19}
Y. Utsumi, {\it Full counting statistics of information content},
\href{https://doi.org/10.1140/epjst/e2018-800043-4}{Eur. Phys. J. Spec. Top. 227, 1911 (2019)}. 

\bibitem{arv-20}
F. Ares, M. A. Rajabpour, J. Viti, {\it Exact full counting statistics for the staggered magnetization and the domain walls in the XY spin chain}, \href{
https://doi.org/10.1103/PhysRevE.103.042107}{Phys. Rev. E \textbf{103}, 042107 (2021)}.


\bibitem{bcp-23}
R. Bonsignori, L. Capizzi, and P. Panopoulos, {\it Boundary Symmetry Breaking in CFT and the String Order Parameter}, \href{https://doi.org/10.1007/JHEP05(2023)027}{JHEP {\bf 05} (2023) 027}.

\bibitem{dvdvdr-23}
G. Del Vecchio Del Vecchio, B. Doyon and P. Ruggiero, \textit{Entanglement
Rényi Entropies from Ballistic Fluctuation Theory: the free fermionic case,}
 \href{
https://doi.org/10.48550/arXiv.2301.02326}{arXiv:2301.02326 (2023).}



\bibitem{Belin-Myers-13-HolChargedEnt}
A. Belin, L.-Y. Hung, A. Maloney, S. Matsuura, R. C. Myers, and T. Sierens, 
{\it Holographic charged R\'enyi entropies}, 
\href{https://doi.org/10.1007/JHEP12(2013)059}{JHEP {\bf 12} (2013) 059}.

 
\bibitem{lr-14}
N. Laflorencie and S. Rachel, 
{\it Spin-resolved entanglement spectroscopy of critical spin chains and Luttinger liquids},
\href{http://dx.doi.org/10.1088/1742-5468/2014/11/P11013}{J. Stat. Mech. (2014) P11013}.


\bibitem{goldstein}
M. Goldstein and E. Sela, 
\textit{Symmetry-Resolved Entanglement in Many-Body Systems},
\href{http://dx.doi.org/10.1103/PhysRevLett.120.200602}{Phys. Rev. Lett. {\bf 120}, 200602 (2018)}.


\bibitem{xavier}
J. C. Xavier, F. C. Alcaraz, and G. Sierra, {\it Equipartition of the entanglement entropy}, \href{https://journals.aps.org/prb/abstract/10.1103/PhysRevB.98.041106}{Phys. Rev. B {\bf  98}, 041106 (2018)}.

\bibitem{amc-22}
F. Ares, S. Murciano, and P. Calabrese, {\it Entanglement asymmetry as a probe of symmetry breaking}, 
\href{https://doi.org/10.1038/s41467-023-37747-8}{Nature Commun. {\bf 14},  2036 (2023)}.

\bibitem{amvc-23}
F. Ares, S. Murciano, E. Vernier, and P. Calabrese, {\it Lack of symmetry restoration after a quantum quench: an entanglement asymmetry study}, \href{https://arxiv.org/abs/2302.03330}{Arxiv:2302.03330}.


\bibitem{cmc-22}
L. Capizzi, S. Murciano and P. Calabrese, {\it R\'enyi entropy and negativity for massless Dirac fermions at conformal interfaces and junctions,} \href{https://doi.org/10.1007/JHEP08(2022)171}{JHEP {\bf 08}, 171 (2022)}.

\bibitem{cmc-22a}
L. Capizzi, S. Murciano and P. Calabrese, {\it R\'enyi entropy and negativity for massless complex boson at conformal interfaces and junctions,} \href{https://doi.org/10.1007/JHEP11(2022)105}{JHEP {\bf 2022}, 105 (2022)}.

\bibitem{cc-04}
P. Calabrese and J. Cardy, 
{\it Entanglement entropy and quantum field theory}, 
\href{http://dx.doi.org/10.1088/1742-5468/2004/06/P06002}{J. Stat. Mech. (2004) P06002}.

\bibitem{cct-12}
P. Calabrese, J. Cardy, and E. Tonni, {\it Entanglement Negativity in Quantum Field Theory}, 
\href{http://dx.doi.org/10.1103/PhysRevLett.109.130502}{Phys. Rev. Lett. {\bf 109}, 130502 (2012)}.

\bibitem{kf-92}
C. L. Kane and M. P. A. Fisher, {\it Transmission through barriers and resonant tunneling in an interacting one-dimensional electron gas},
\href{https://doi.org/10.1103/PhysRevB.46.15233}{Phys. Rev. B {\bf 46}, 15233 (1992)}.

\bibitem{ss-08}
K. Sakai and Y. Satoh, \textit{Entanglement through conformal interfaces}, \href{https://doi.org/10.1088/1126-6708/2008/12/001}{JHEP 12 (2008) 001}.

\bibitem{bb-15}
E. Brehm and I. Brunner, \textit{Entanglement entropy through conformal interfaces in the 2D Ising model},
\href{https://doi.org/10.1007/JHEP09(2015)080}{JHEP 09 (2015) 80}.

\bibitem{bbr-13}
C. Bachas, I. Brunner, and D. Roggenkamp, \textit{Fusion of Critical Defect Lines in the 2D Ising Model}, 
\href{https://doi.org/10.1088/1742-5468/2013/08/P08008}{J. Stat. Mech. (2013) P08008}.

\bibitem{tm-21}
M. Mintchev and E. Tonni, {\it Modular Hamiltonians for the massless Dirac field in the presence of a defect}, 
\href{https://link.springer.com/article/10.1007/JHEP03(2021)205}{JHEP 03 (2021) 205}.


\bibitem{p-05}
I. Peschel,
{\it Entanglement entropy with interface defects}, \href{https://doi.org/10.1088/0305-4470/38/20/002}{J. Phys. A {\bf 38}, 4327 (2005)}.

\bibitem{Levine}
G. Levine, {\it Entanglement entropy in a boundary impurity model}, \href{
https://doi.org/10.1103/PhysRevLett.93.266402}{Phys. Rev. Lett. {\bf 93}, 266402 (2004).}

\bibitem{ep-10}
V. Eisler and I. Peschel,
{\it Solution of the fermionic entanglement problem with interface defects}, 
\href{https://doi.org/10.1002/andp.201000055}{Ann. Phys. (Berlin) {\bf 522}, 679 (2010)}.

\bibitem{ep-12}
I. Peschel and V. Eisler,
{\it Exact results for the entanglement across defects in critical chains}, 
\href{https://doi.org/10.1088/1751-8113/45/15/155301}{J. Phys. A {\bf 45}, 155301 (2012)}.

\bibitem{ep2-12}
V. Eisler and I. Peschel, \textit{On entanglement evolution across defects in critical chains}, \href{https://doi.org/10.1209/0295-5075/99/20001}{EPL {\bf 99}, 20001 (2012)}.

\bibitem{ce-22}
L. Capizzi and V. Eisler, \textit{Entanglement evolution after a global quench across a conformal defect}, 
\href{https://doi.org/10.21468/SciPostPhys.14.4.070}{SciPost Phys. {\bf 14}, 070 (2023)}.




\bibitem{csrc-23}
L. Capizzi, S. Scopa, F. Rottoli, and P. Calabrese, \textit{Domain wall melting across a defect}, \href{https://doi.org/10.1209/0295-5075/acb50a}{EPL \textbf{141}, 31002 (2023)}.




\bibitem{cmv-12}
P. Calabrese, M. Mintchev, and E. Vicari, \textit{Entanglement Entropy of Quantum Wire Junctions}, 
\href{https://doi.org/10.1088/1751-8113/45/10/105206}{J. Phys. A {\bf 45}, 105206 (2012)}.

\bibitem{rs-22}
A. Roy and H. Saleur, {\it Entanglement entropy in critical quantum spin chains with boundaries and defects},
\href{http://dx.doi.org/10.1007/978-3-031-03998-0_3}{In: Bayat, A., Bose, S., Johannesson, H. (eds) Entanglement in Spin Chains. Quantum Science and Technology. Springer, Cham (2022)}.

\bibitem{sths-22}
H. Schlömer, C. Tan, S. Haas, and H. Saleur, {\it Parity effects and universal terms of O(1) in the entanglement near a boundary},
\href{http://dx.doi.org/10.21468/SciPostPhys.13.5.110}{SciPost Phys. 13, 110 (2022)}.


\bibitem{bm-06}
B. Bellazzini and M. Mintchev, {\it Quantum Fields on Star Graphs}, 
\href{https://doi.org/10.1088/0305-4470/39/35/011}{J. Phys. A {\bf 39}, 11101  (2006)}.

\bibitem{bms-07}
B. Bellazzini, M. Mintchev and P. Sorba, {\it Bosonization and Scale Invariance on Quantum Wires}, 
\href{https://doi.org/10.1088/1751-8113/40/10/017}{J. Phys. A {\bf 40}, 2485 (2007)}.

\bibitem{cmv-11}
P. Calabrese, M. Mintchev, and E. Vicari,
{\it The entanglement entropy of one-dimensional gases},
\href{http://dx.doi.org/10.1103/PhysRevLett.107.020601/}{Phys. Rev. Lett. {\bf 107}, 020601 (2011)}.

\bibitem{cmv-12a}
P. Calabrese, M. Mintchev, and E. Vicari,
{\it Exact relations between particle fluctuations and entanglement in Fermi gases}, 
\href{http://dx.doi.org/10.1209/0295-5075/98/20003}{EPL {\bf 98}, 20003 (2012)}.

\bibitem{riccarda}
R. Bonsignori, P. Ruggiero, and P. Calabrese, 
{\it Symmetry resolved entanglement in free fermionic systems}, 
\href{https://doi.org/10.1088/1751-8121/ab4b77}{J. Phys. A {\bf 52}, 475302 (2019)}.




\bibitem{mgc-20}
S. Murciano, G. Di Giulio, and P. Calabrese, 
{\it Entanglement and symmetry resolution in two dimensional free quantum field theories}, 
\href{https://link.springer.com/article/10.1007/JHEP08(2020)073}{JHEP {\bf 08} (2020) 073}.

\bibitem{fmc-22}
A. Foligno, S. Murciano, P. Calabrese, 
\textit{Entanglement resolution of free Dirac fermions on a torus}, \href{https://doi.org/10.1007/JHEP03(2023)096}{JHEP 03 (2023) 096}.

\bibitem{vek-21}
V. Vitale, A. Elben, R. Kueng, A. Neven, J. Carrasco, B. Kraus, P. Zoller, P. Calabrese, B. Vermersch, and M. Dalmonte, 
{\it Symmetry-resolved dynamical purification in synthetic quantum matter}, 
\href{https://doi.org/10.21468/SciPostPhys.12.3.106}{SciPost Phys. {\bf 12}, 106 (2022)}.

\bibitem{ncv-21}
A. Neven, J. Carrasco, V. Vitale, C. Kokail, A. Elben, M. Dalmonte, P. Calabrese, P. Zoller, B. Vermersch, R. Kueng, and B. Kraus,
{\it Symmetry-resolved entanglement detection using partial transpose moments},
\href{https://doi.org/10.1038/s41534-021-00487-y}{npj Quantum Inf. {\bf 7}, 152 (2021)}.

\bibitem{rvm-22}
A. Rath, V. Vitale, S. Murciano, M. Votto, J. Dubail, R. Kueng, C. Branciard, P. Calabrese, B. Vermersch, {\it Entanglement barrier and its symmetry resolution: theory and experiment}, \href{https://doi.org/10.1103/PRXQuantum.4.010318}{PRX Quantum {\bf 4}, 010318 (2023)}.

\bibitem{longrange}
 F. Ares, S. Murciano and P. Calabrese, 
{\it Symmetry-resolved entanglement in a long-range free-fermion chain},
\href{
https://doi.org/10.1088/1742-5468/ac7644}{J. Stat. Mech. (2022) 063104}.

\bibitem{cct-13}
P. Calabrese, J. Cardy, and E. Tonni, {\it Entanglement negativity in extended systems: a field theoretical approach},
\href{http://dx.doi.org/10.1088/1742-5468/2013/02/P02008}{J. Stat. Mech. P02008 (2013)}.

\bibitem{altland}
A. Altland and B. Simons, 
{\it Condensed Matter Field Theory}, 
Cambridge University Press, 2nd Edition (2010).
\bibitem{mgc-gapped}
S. Murciano, G. Di Giulio and P. Calabrese, 
{\it Symmetry resolved entanglement in gapped integrable systems: a corner transfer matrix approach}, 
\href{https://dx.doi.org/10.21468/SciPostPhys.8.3.046}{SciPost Phys. {\bf 8}, 046 (2020)}.

\bibitem{hcc-21}
D. X. Horvath, L. Capizzi, and P. Calabrese,
{\it U(1) symmetry resolved entanglement in free 1+1 dimensional field theories via form factor bootstrap},
\href{https://doi.org/10.1007/JHEP05(2021)197}{JHEP {\bf 05} (2021) 197}.

\bibitem{goldstein1}
E. Cornfeld, M. Goldstein, and E. Sela, 
{\it Imbalance Entanglement: Symmetry Decomposition of Negativity}, 
\href{http://dx.doi.org/10.1103/PhysRevA.98.032302} {Phys. Rev. A {\bf 98}, 032302 (2018)}.

\bibitem{mbc-21}
S. Murciano, R. Bonsignori and P. Calabrese, 
{\it Symmetry decomposition of negativity of massless free fermions},
\href{https://doi.org/10.21468/SciPostPhys.10.5.111}{SciPost Phys. {\bf 10}, 111 (2021)}.




\bibitem{spectrum}
P. Calabrese, A. Lefevre, {\it 
Entanglement spectrum in one-dimensional systems},
\href{https://doi.org/10.1103/PhysRevA.78.032329
}{Phys. Rev A {\bf 78}, 032329 (2008).}

\bibitem{mt2-21}
M. Mintchev and E. Tonni, {\it  Modular hamiltonians for the massless dirac field in the presence of
  a defect,} \href{https://link.springer.com/article/10.1007/JHEP03(2021)205}{JHEP {\bf 03} (2021) 205.}

\end{thebibliography}
 \end{document}